\documentclass[aps,prb,twocolumn,nofootinbib,longbibliography,superscriptaddress]{revtex4-2}
\usepackage[dvipdfmx]{graphicx}
\usepackage{amsfonts}
\usepackage{amscd}
\usepackage{amsmath}
\usepackage{amssymb}
\usepackage{newtxtext,newtxmath}
\usepackage{here}
\usepackage{tabularx}
\usepackage{etoolbox}
\usepackage{bm}
\usepackage{mathrsfs}
\usepackage{listings}
\usepackage{color}
\usepackage{braket}
\usepackage[normalem]{ulem}  
\usepackage{hyperref}

\newcommand{\beq}{\begin{eqnarray}}
\newcommand{\eeq}{\end{eqnarray}}

\newcommand\erase{\bgroup\markoverwith{\textcolor{blue}{\rule[.5ex]{2pt}{1pt}}}\ULon}

\begin{document}
\title{Spin caloritronics as a probe of nonunitary superconductors}

\author{Taiki Matsushita$^\ast$}
\affiliation{Department of Physics, Graduate School of Science, Kyoto University, Kyoto 606-8502, Japan}
\affiliation{Center for Gravitational Physics and Quantum Information, Yukawa Institute for Theoretical Physics, Kyoto University, Kyoto 606-8502, Japan}
\author{Takeshi Mizushima}
\affiliation{Department of Materials Engineering Science, Osaka University, Toyonaka, Osaka 560-8531, Japan}
\author{Yusuke Masaki}
\affiliation{Department of Applied Physics, Graduate School of Engineering, Tohoku University, Sendai, Miyagi 980-8579, Japan}
\author{Satoshi Fujimoto}
\affiliation{Department of Materials Engineering Science, Osaka University, Toyonaka, Osaka 560-8531, Japan}
\affiliation{Center for Quantum Information and Quantum Biology, Osaka University, Toyonaka, Osaka 560-8531, Japan}
\author{Ilya Vekhter}
\affiliation{Department of Physics and Astronomy, Louisiana State University, Baton Rouge, LA 70803-4001, USA}

\date{\today}
\begin{abstract}
{\bf Superconducting spintronics explores the interplay between superconductivity and magnetism, sparking significant interest in nonunitary superconductors as a platform for novel magneto-superconducting phenomena. 
However, identifying nonunitary superconductors remains challenging. 
We demonstrate that spin current driven by thermal gradients sensitively probes the nature of the condensate in nonunitary superconductors. 
Spin polarization of the condensate in momentum space induces the superconducting spin Seebeck effect, where a spin current is generated along thermal gradients without a thermoelectric charge current. Notably, the nonvanishing superconducting spin Seebeck effect provides a smoking gun evidence of nonunitary superconductivity because it reflects the spin polarization of the condensate in momentum space, irrespective of whether the net pair spin magnetization vanishes. At the same time, the spin-chirality of the condensate induces the spin-Nernst effect, where a spin current is generated perpendicular to thermal gradients in nonunitary superconductors. These spin caloritronics phenomena offer a definitive probe of nonunitary superconductors.}

\vspace{5mm}
{$^*$Corresponding author. Email: matsushita.taiki.8j@kyoto-u.ac.jp}

\vspace{5mm}
{\bf Teaser:}
{Superconducting spin Seebeck effect is proposed as probes for nonunitary superconductors.}
\end{abstract}

\maketitle


\section*{Introduction}
Classification of superconducting states and identification of unusual superconductivity underlie both the efforts to understand mechanisms of electron pairing and applications of superconducting materials. At the most fundamental level, determining whether a superconducting state breaks the time-reversal and/or point group symmetries of the host crystal lattice restricts theories of the possible electron pairing in a given family of materials~\cite{sigrist1991phenomenological,annettreview_1990, Volovik1985a}. More recently, with the emergence of topological electronic matter as a potential driver for applications, 
reliable methods to identify topological superconductivity have been in demand~\cite{qi2011topological,sato2016majorana,sato2017topological}. Similarly, proposals to utilize the interplay between superconductivity and magnetism for superconducting spintronics~\cite{linder2015superconducting,eschrig2015spin,sharma2023spin,bathen2017spin}, have brought a renewed focus on the studies of the spin properties of the paired electron states.

The superconducting order parameter matrix in spin space, $\Delta(\bm k)=\left[\psi(\bm k)+\bm d(\bm k)\cdot\bm\sigma\right]\left(i\sigma_y\right)
$, where $\bm \sigma=(\sigma_x,\sigma_y,\sigma_z)$ is a vector of the Pauli matrices, has a scalar spin-singlet ($\psi(\bm k)$) and vector spin-triplet ($\bm d(\bm k)$) components at each momentum $\bm k$ on the Fermi surface. 
The intrinsic spin polarization of the condensate in momentum space, ${\bm S}(\bm k)={\rm Tr}\left[\Delta^\dagger(\bm k) \bm\sigma \Delta(\bm k)\right]=-i{\bm d}^\ast(\bm k)\times {\bm d}(\bm k)$, is inherent in nonunitary superconductors (NUSCs), for which ${\bm d}^\ast(\bm k)\times {\bm d}(\bm k)\neq 0$. 
However, even if Cooper pairs at each $\bm k$ have a well-defined spin polarization, their net pair spin magnetization averaged over the entire Fermi surface may still vanish. Consequently, the clear identification of NUSCs remains an outstanding problem. We show below that the spin current response to thermal gradients in superconductors provides unique information about the nontrivial structures of the order parameter.

\begin{figure}[t]
\includegraphics[width=85mm]{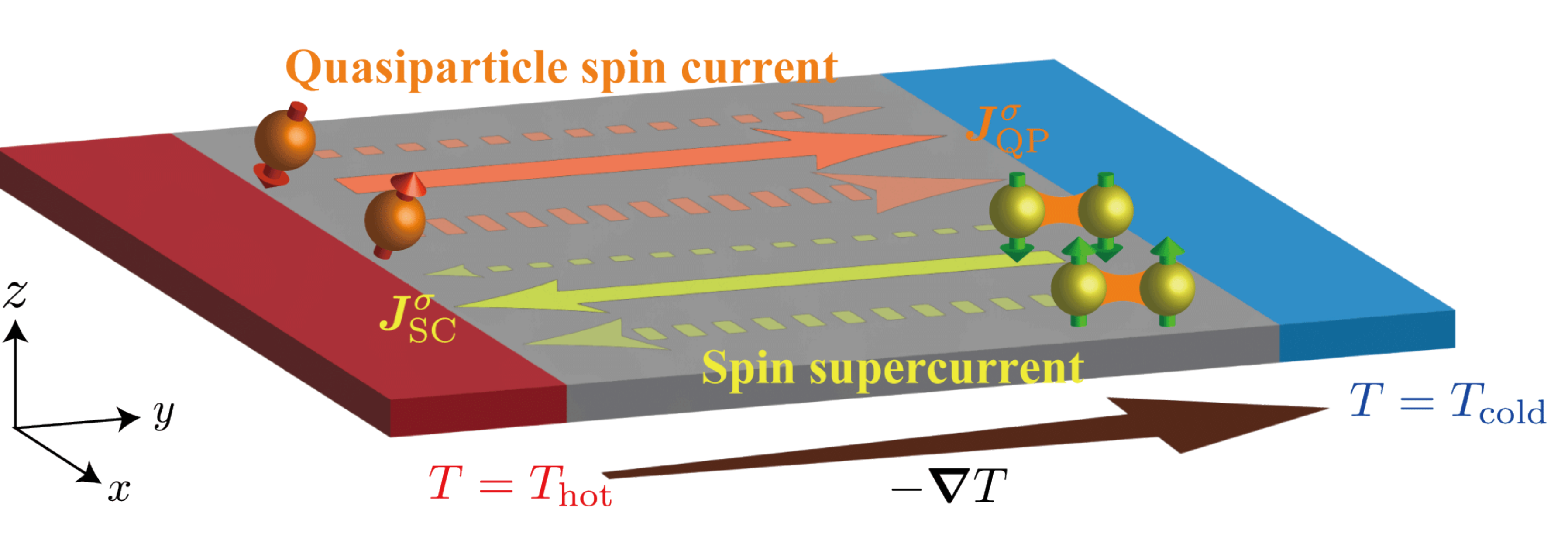}
\caption{{\bf Sketch of the physical origin of the superconducting spin Seebeck effect in nonunitary superconductors.} 
The nonunitary superconductor {(grey) is in contact with a heater (red) and heat sink (blue), 
with the temperatures $T=T_{\rm hot}(T_{\rm cold})$ respectively}.
The orange (yellow) arrows show the spin current carried by quasiparticles (Cooper pairs), and the broken arrows show the spin-polarized currents.
The spin supercurrent is generated to cancel the dissipative electric current carried by quasiparticles (see text for details).
The brown arrow represents thermal gradients.
}
\label{fig: setup}
\end{figure}
 
We propose a new type of spin transport phenomenon unique to superconductors, the superconducting spin Seebeck effect (SSSE).
The SSSE designation indicates the generation of net spin current along the thermal gradient in the absence of thermoelectric charge current. The thermoelectric charge current is fully canceled by the supercurrent in bulk superconductors, as shown in Fig.~\ref{fig: setup}.
The SSSE should therefore be distinguished from the spin-Seebeck effect for magnons in magnetic systems and the spin-dependent Seebeck effect for spin-polarized electron systems, because the electric voltage commonly observed in the latter case is absent in bulk superconductors~\cite{bauer2012spin,uchida2021transport,uchida2008observation}.
The SSSE provides a smoking gun signature of NUSCs, irrespective of whether the net spin magnetization of nonunitary Cooper pairs vanishes or not.
We also emphasize that the spin-Nernst effect (SNE), spin current generation perpendicular to the thermal gradient~\cite{uchida2021transport,meyer2017observation,sheng2017,bose2018}, is a probe of the spin-chirality of the condensate~\cite{matsushita2022spin}. Our work demonstrates that spin current is uniquely suited for examining the coupling between the relative orbital motion (the $\bm k$-dependence of the $d$-vector) and the spin (the direction of $d$-vector) of spin-triplet Cooper pairs.

We carry out our calculations in the framework of quasiclassical nonequilibrium Keldysh-Eilenberger theory~\cite{eilenberger1968transformation,serene1983quasiclassical}.
The key point of our analysis is to properly account for the absence of the thermoelectric charge current in bulk superconductors~\cite{ginzburg1978thermoelectric,Shelly2016}.
As shown in Fig.~\ref{fig: setup} and discussed below, the thermal gradient induces a spatial change in the U(1) phase of the superconducting order parameter, leading to a counterflow of supercurrent that cancels out the dissipative thermoelectric charge current carried by quasiparticles. Importantly, this cancellation permits the existence of the thermoelectric spin current in NUSCs.

\begin{figure*}[t]
\includegraphics[width=140mm]{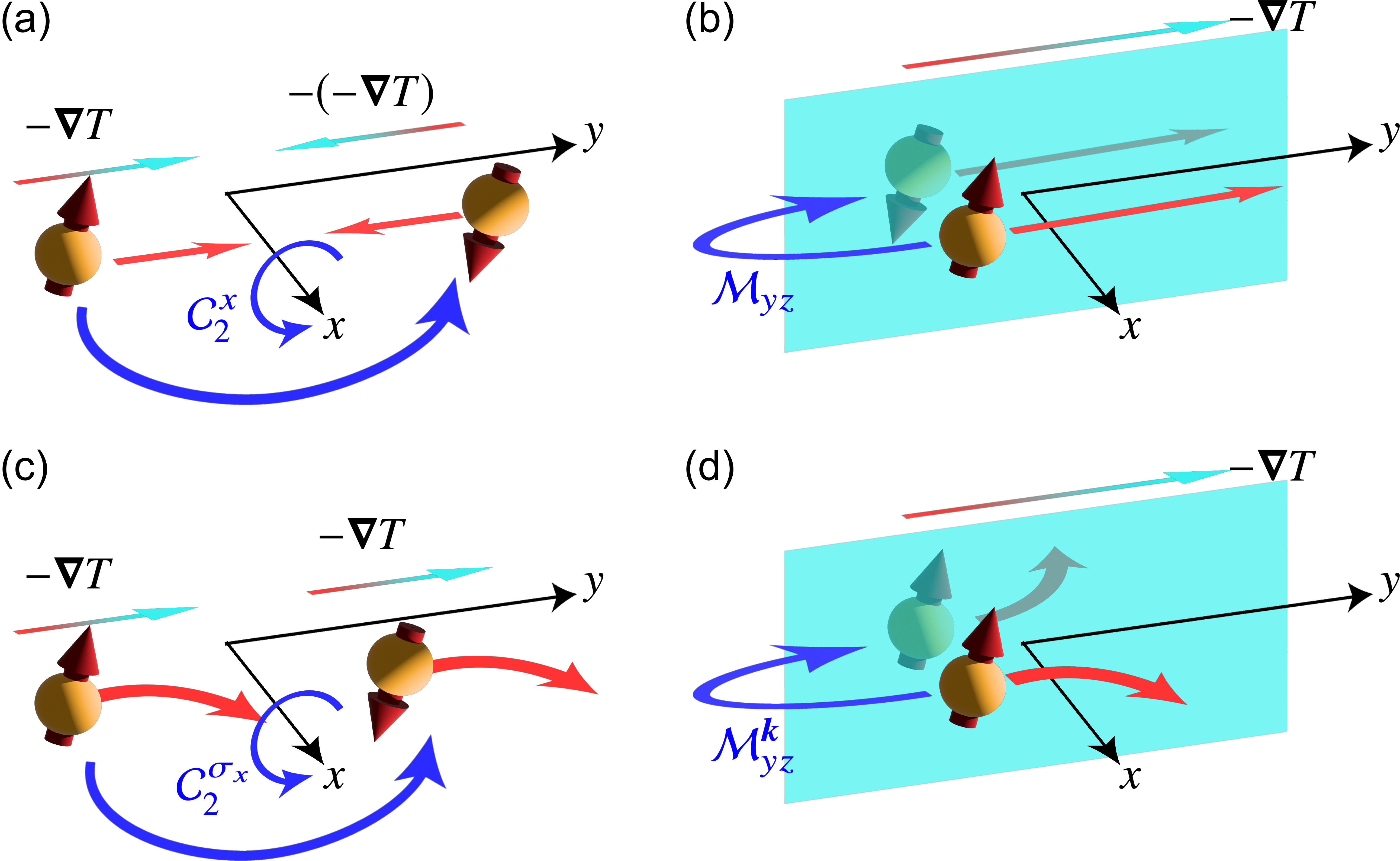}
\caption{
{\bf Symmetry operations and their constraints on the spin transport.} (a) the two-fold rotation along the $x$-direction ($\mathcal{C}_2^x$), (b) the mirror reflection in the $yz$ plane ($\mathcal{M}_{yz}$), (c) the two-fold spin rotation along the $x$-direction ($\mathcal{C}_2^{\sigma_x}$), and (d) the momentum mirror reflection in the $yz$ plane ($\mathcal{M}^{\bm k}_{yz}$).}
\label{fig: symmetry}
\end{figure*}

Our results are relevant to a wide range of superconductors of intense current interest. 
While the early belief was that superconductivity is not compatible with magnetic moments because they break the Cooper pairs~\cite{tinkham}, currently there are multiple candidates for nonunitary superconductivity.
The nonunitary superfluid state is established in $^3$He under applied magnetic fields~\cite{vollhardt2013superfluid}, and nonunitary superconductivity has been proposed in several
uranium compounds, such as
U$_{1-x}$Th$_x$Be$_{13}$ and UTe$_2$.
U$_{1-x}$Th$_x$Be$_{13}$ shows a double superconducting phase transition with decreasing temperature for $0.19\leq x \leq 0.40$~\cite{ott1985phase, balatsky2006impurity,shimizu2017quasiparticle}, and its low-temperature phase is strongly suggested to be nonunitary~\cite{PhysRevLett.65.2816,machida2018spin,mizushima2018topology}.
While more controversial, some experiments indicate the possible realization of nonunitary superconductivity in UTe$_2$~\cite{jiao2020chiral,hayes2021multicomponent,ishihara2023chiral}.
Ferromagnetic superconductors, such as UCoGe, URhGe, and UGe$_2$, where 
superconductivity emerges from the ferromagnetic metallic phase~\cite{aoki2019review},
are also candidates for NUSCs according to the group-theoretical analysis ~\cite{mineev2002superconducting}. 
The nonunitary pairing proposed in Ref~\cite{mineev2002superconducting} successfully explains the angle-resolved NMR experiment in UCoGe~\cite{hattori2012superconductivity}.
As the number of candidate NUSCs increases, we expect that our results will become relevant to a wider and wider range of compounds.

This paper is organized as follows.
We first define the corresponding transport conductivities and identify the symmetry that must be broken for their nonvanishing values. 
To aid this, we define subsidiary superconducting orders, i.e., the averages of the orbital and spin angular momenta of the condensate that allow SSSE and SNE. 
We next discuss the calculated results for simple models of NUSCs with/without the net pair spin magnetization, clarifying that the SSSE is generic in NUSCs.
The Methods section provides the technical details of our calculations. 
Because the spin current cannot be directly measured, we propose an experimental setup for observing these signals in NUSCs, and estimate their magnitude, showing that it is well within the measurable range.
Finally, we describe the relevance of our calculated results to realistic materials and future prospects.

\section*{Results}
\subsection*{Symmetry constraints on
thermospin conductivities.}
We define the transport coefficients for the SSSE and the SNE and derive the symmetry constraints under which they do not vanish.
First, we re-emphasize the physics behind the vanishing of the thermoelectric charge current in the bulk of superconductors.
In the Meissner state, the magnetic field is excluded from the bulk according to the London equation, $\nabla^2\bm B=\bm B/\lambda^2$, where $\lambda$ is the magnetic penetration depth.
This exclusion suppresses the thermoelectric response~\cite{ginzburg1978thermoelectric,Shelly2016} because the magnetic field is related to the electric current via the Maxwell-Amp\`{e}re law, $\bm J_{\rm QP}+\bm J_{\rm SC}={\bm \nabla}\times \bm B$, where $\bm J_{\rm QP}$ and $\bm J_{\rm SC}$ are the dissipative thermoelectric current carried by quasiparticles and the supercurrent, respectively.
When the Meissner effect is complete, $\bm B=0$, the thermoelectric charge current vanishes, $\bm J_{\rm QP}+\bm J_{\rm SC}=0$.
Hence, in the superconducting state, we should compute the spin current in the absence of the thermoelectric charge current.

Let us define the thermospin conductivity tensor as,
\beq
\alpha^{\sigma_{\rho}}_{\mu \nu} \equiv J^{\sigma_{\rho}}_\mu/(-\partial_\nu T), 
\eeq
where $J^{\sigma_{\rho}}_\mu$ is the spin current with polarization along the $\rho$ axis flowing in the $\mu$ direction in response to the thermal gradient along the $\nu$ axis. Each of the indices ($\rho,\mu$ and $\nu$) denotes the cartesian coordinates ($x,y$ and $z$).
The diagonal tensor elements, $\alpha^{\sigma_{\rho}}_{\mu \mu}$, describe the SSSE and are referred to as the spin Seebeck conductivities (SSCs) while the off-diagonal elements, $\alpha^{\sigma_{\rho}}_{\mu \nu}\;(\mu\neq \nu)$, are for the SNE and are called the spin Nernst conductivities (SNCs).

Nonvanishing SSC and SNC require the breaking of discrete symmetries of the underlying lattice.
Let $\mathcal{M}_{\mu \nu}$, $\mathcal{C}_2^{\mu}$, $\mathcal{I}$, and $\mathcal{T}$ be the operations of the mirror reflection in the $\mu \nu$ plane, the two-fold rotation about the $\mu$ axis, the spatial inversion, and the time-reversal, respectively. 
Two-fold rotational symmetry restricts possible components of the SSSE.
As depicted in Fig.~\ref{fig: symmetry} (a), if the system is invariant with respect to the two-fold rotation along the $x$-direction ($\mathcal{C}_2^{x}$), it immediately 
requires $\alpha^{\sigma_y}_{\mu \mu}=\alpha^{\sigma_z}_{\mu \mu}=0$, while the SSC with the spin polarization along the $x$ axis ($\alpha^{\sigma_x}_{\mu \mu}$) is allowed.
The mirror reflection symmetry restricts the SSCs in the same way because it is defined as the combination of the two-fold rotation and the spatial inversion~\cite{Seemann2015}.
For instance, the mirror reflection symmetry in the $yz$ plane ($\mathcal{M}_{yz}$) requires $\alpha^{\sigma_y}_{\mu \mu}=\alpha^{\sigma_z}_{\mu \mu}=0$ because of $\mathcal{C}_2^{x}=\mathcal{I}\mathcal{M}_{yz}$ (see Fig.~\ref{fig: symmetry} (b)).
Hence, all SSCs become zero in the presence of the two-fold rotational (or mirror reflection) symmetries for more than one crystalline axis (mirror plane).
Note that time-reversal symmetry ($\mathcal{T})$ by itself does not prohibit the SSSE, 
even though the time-reversal symmetry is broken in many cases with finite SSCs~\cite{Seemann2015}.

The SNE is also restricted by the two-fold rotational and mirror reflection symmetries. For instance, the two-fold rotational symmetry with respect to the $x$ axis ($\mathcal{C}_2^{x}$) requires,
\beq
\alpha^{\sigma_x}_{xy}=\alpha^{\sigma_x}_{yx}=\alpha^{\sigma_x}_{zx}=\alpha^{\sigma_x}_{xz}=\alpha^{\sigma_y}_{yz}=\alpha^{\sigma_y}_{zy}=\alpha^{\sigma_z}_{yz}=\alpha^{\sigma_z}_{zy}=0.\nonumber\\
\eeq
The SNCs are subject to similar restrictions from the mirror reflection symmetry in the $yz$ plane ($\mathcal{M}_{yz}$), in analogy with the above discussion of the SSSE~\cite{Seemann2015}.

The point group symmetries, such as the two-fold rotational or mirror reflection symmetries, do not prohibit the tensor elements $\alpha^{\sigma_{\rho}}_{\mu \nu}$ with $(\mu, \nu, \rho) = (x, y, z)$ and its permutations.
Further constraints can be obtained if there is a two-fold spin rotational symmetry, i.e., a $\pi$-rotation of spin which keeps the crystal momentum unchanged.
The symmetry for this two-fold spin rotation naturally appears in the weak spin-orbit coupling (SOC) limit, where the spin and the momentum can be rotated independently. 
However, as we see below, even in the presence of the SOC, once the $d$-vector is fixed by SOC, there may remain the two-fold spin rotational symmetry around a particular axis.
If this symmetry holds, consider two-fold spin rotation ($\mathcal{C}^{\sigma_x}_2$) about the $x$ axis, which transforms the spin as $(\sigma_x,\sigma_y,\sigma_z)\to (\sigma_x,-\sigma_y,-\sigma_z)$ while $\bm k$ remains constant. This emergent symmetry for spin ensures equal quasiparticle flow in each of the opposite spin sectors and, hence, $\alpha^{\sigma_y}_{\mu \nu}=\alpha^{\sigma_z}_{\mu \nu}=0$ for all $\mu,\nu=x,y,z$ (see Fig.~\ref{fig: symmetry} (c)).
Under the same assumptions, we can define the momentum mirror reflection as $\mathcal{M}^{\bm k}_{yz}=\mathcal{M}_{yz}\mathcal{C}^{\sigma_x}_2$, which preserves the spin orientation but reverses the transverse momentum, $k_x\rightarrow -k_x$.
This operation ensures an equal population of quasiparticles with $k_x$ and $-k_x$ in each spin sector, resulting in $\alpha_{xy}^{\sigma_x}=\alpha_{xy}^{\sigma_y}=\alpha_{xy}^{\sigma_z}=0$ (see Fig.~\ref{fig: symmetry} (d)).

\begin{table*}
  \centering
  \begin{tabular}{c|c|c}
    \hline
    Subsidiary order & Breaking symmetries & Candidate materials \\
    \hline \hline
    \begin{tabular}{c}
    Chirality\\
    $\langle L_\mu\rangle$
    \end{tabular}
    &$\mathcal{T}$, $\mathcal{M}_{\nu \rho}$, $\mathcal{C}^\nu_{2}$
    & 
    \begin{tabular}{l}
    $^3$He (A, A$_1$$^\circ$ and A$_2$$^\circ$ phases)~\cite{vollhardt2013superfluid}, Sr$_2$RuO$_4$~\cite{kallin2012chiral}, URu$_2$Si$_2$~\cite{kittaka2016evidence,yamashita2015colossal,PhysRevB.90.184518},
     \\SrPtAs~\cite{PhysRevB.87.180503,PhysRevB.99.144510}, UPt$_3$ (B phase)~\cite{schemm2014observation}, U$_{1-x}$Th$_x$Be$_{13}$ (B$^\circ$-phase)~\cite{PhysRevLett.65.2816,mizushima2018topology,machida2018spin,PhysRevB.39.2200}, \\UCoGe$^\circ$~\cite{mineev2002superconducting,hattori2012superconductivity}, URhGe$^\circ$~\cite{mineev2002superconducting}, UGe$_2$$^\circ$~\cite{mineev2002superconducting}
    \end{tabular}\\
    \hline
    \begin{tabular}{c}
    Spin-chirality\\
    $\langle L_\mu S_\mu\rangle$
    \end{tabular}
    &$\mathcal{M}^{\bm k}_{\nu \rho},\; \mathcal{C}_{2}^{\sigma_\nu}$
    &
    \begin{tabular}{l}
    $^3$He (B, A$_1$$^\circ$, and A$_2$$^\circ$ phases)~\cite{vollhardt2013superfluid}, Cu$_x$Bi$_2$Se$_3$~\cite{yonezawa2018nematic},
    \\U$_{1-x}$Th$_x$Be$_{13}$ (A, B$^\circ$ and C phase)~\cite{mizushima2018topology,machida2018spin,PhysRevB.96.100505,PhysRevB.39.2200}, UCoGe$^\circ$~\cite{mineev2002superconducting,hattori2012superconductivity},\\
    URhGe$^\circ$~\cite{mineev2002superconducting}, UGe$_2$$^\circ$~\cite{mineev2002superconducting},UTe$_2$~\cite{ran2019nearly,knebel2019field,PhysRevLett.123.217001,PhysRevB.103.094504},
    URhGe$^\circ$~\cite{mineev2002superconducting}, UGe$_2$$^\circ$~\cite{mineev2002superconducting}\\
    \end{tabular}\\
    \hline
    \begin{tabular}{c}
    Pair spin magnetization\\
    $\langle S_\mu\rangle$
    \end{tabular}
    & $\mathcal{T}$, $\mathcal{C}^\nu_{2}$,\;$\mathcal{C}_{2}^{\sigma_\nu}$&
    \begin{tabular}{l}
    $^3$He (A$_1$$^\circ$and A$_2$$^\circ$ phase)~\cite{vollhardt2013superfluid}, U$_{1-x}$Th$_x$Be$_{13}$$^\circ$ (B$^\circ$-phase)~\cite{PhysRevLett.65.2816,mizushima2018topology,machida2018spin,PhysRevB.39.2200},
    \\UCoGe$^\circ$~\cite{mineev2002superconducting,hattori2012superconductivity}, URhGe$^\circ$~\cite{mineev2002superconducting}, UGe$_2$$^\circ$~\cite{mineev2002superconducting} \\
    \end{tabular}\\
    \hline
  \end{tabular}
  \caption{{\bf Subsidiary orders, the corresponding broken symmetries, and the candidate materials.} $\circ$ represents the nonunitary superconducting phase. $\mu, \nu$, and $\rho$ are orthogonal crystalline axes.
  }
  \label{subsidiaryorder}
\end{table*}

\subsection*{Subsidiary classification of superconducting orders allowing SSSE/SNE}
It is helpful to define subsidiary orders that indicate the breaking of symmetries that otherwise prohibit the SSSE and the SNE.
A general superconducting order parameter matrix in spin space 
with the spin quantization axis along the $z$-direction is given by,
\beq
\label{scmat}
\Delta(\bm k)&=&
\begin{pmatrix}
\Delta_{\uparrow \uparrow}(\bm k)&&\Delta_{\uparrow \downarrow}(\bm k)\\ \Delta_{\downarrow \uparrow}(\bm k)&&\Delta_{\downarrow \downarrow}(\bm k)
\end{pmatrix}\nonumber\\
&=&
\begin{pmatrix}
-d_x(\bm k)+i d_y(\bm k)&&d_z(\bm k)+\psi(\bm k)\\
d_z(\bm k)-\psi(\bm k)&&d_x(\bm k)+i d_y(\bm k)
\end{pmatrix},
\eeq
where $\psi(\bm k)$ and ${\bm d}(\bm k)$ are the scalar spin-singlet and vector spin-triplet order parameters, respectively.
For spin-triplet Cooper pairs, we define the chirality, $\braket{\bm L}$, the spin-chirality, $\braket{S_\mu L_\mu}$, and the pair spin magnetization, $\braket{\bm S}$, as~\cite{mineev2002superconducting},
\beq
\label{chirality}
\braket{\bm L}&=&-i \sum_{\mu}\bigg\langle \hat{d}_\mu^\ast \left(\bm k_{\rm F} \times \frac{\partial}{\partial \bm k_{\rm F}} \hat{d}_\mu \right) \bigg\rangle_{\rm FS},\\
\label{spinchirality}
\braket{S_\mu L_\mu}&=&-\frac{i}{2}\sum_{\sigma_\mu=\pm1}\sigma_\mu \bigg\langle \hat{\chi}_{\sigma_\mu}^\ast \bigg({\bm k}_{\rm F}  \times \frac{\partial}{\partial \bm k_{\rm F}} \hat{\chi}_{\sigma_\mu}\bigg)_{\mu} \bigg\rangle_{\rm FS},\\
\label{spinmagnetization}
\braket{\bm S}&=&-i \big\langle \hat{\bm d}^\ast \times \hat{\bm d}\big\rangle_{\rm FS},
\eeq
where the indices $\mu,\nu=x,y,z$ represent the vector components,
$\bm k_{\rm F}$ is the Fermi momentum, and $\langle \cdots \rangle_{\rm FS}$ is the normalized Fermi surface average (so that $\langle 1 \rangle_{\rm FS}=1$).
Moreover, $\hat{\bm d}(\bm k_{\rm F})= \bm d(\bm k_{\rm F})/ \sqrt{\langle |\bm d(\bm k_{\rm F})|^2\rangle_{\rm FS}}$ so that they are independent of the amplitude and the global U(1) phase of the superconducting gap.
To define the spin-chirality, we also introduced the notation $\hat{\chi}_{\sigma_\mu}(\bm k_{\rm F})$ for the gap function for the Cooper pairs with spin $\sigma_\mu=\pm 1$. The explicit form is given by $\hat{\chi}_{\sigma_{\mu}}(\bm k_{\rm F}) = - \sigma_{\mu} \hat{d}_{\nu}(\bm k_{\rm F}) + i \hat{d}_{\rho}(\bm k_{\rm F})$
for $(\mu, \nu, \rho) = (x, y, z), (y, z, x)$, and $(z,x,y)$.

The spin-chirality and the pair spin magnetization are intrinsic to spinful condensates and limited to spin-triplet superconductors, while the chirality can be defined both in spin-singlet and spin-triplet superconductors, 
as the integral of the orbital angular momentum of the condensate, $\bm L$, in momentum space.
For spin-singlet superconductors in Eq.~\eqref{chirality}, $\hat{\bm{d}}(\bm{k}_{\rm F})$ should be replaced by $\hat{\psi}(\bm k_{\rm F})= \psi(\bm k_{\rm F})/ \sqrt{\langle |\psi(\bm k_{\rm F})|^2\rangle_{\rm FS}}$.
Nonvanishing chirality is generated by the complex momentum dependence of the order parameters, such as $\hat{\bm d}(\bm k_{\rm F}),\; \hat{\psi}(\bm k_{\rm F}) \propto (k_x+ik_y)^n\; (n\in \mathbb{Z},\; |n|\geq 1)$~\cite{kallin2016chiral,ghosh2020recent}.
The spin-chirality denotes the difference in the chirality between the spin sectors.
It generally appears when the $d$-vector has multiple components and changes its direction in spin space along the Fermi surface.
One example is a helical superconducting order with $\hat{\bm d}(\bm k_{\rm F}) \propto (k_x, k_y, 0)$.
This order parameter describes Cooper pairs with $L_z=\pm 1$ and $S_z=\mp 1$, yielding nonvanishing spin-chirality, $\braket{S_z L_z}$~\cite{matsushita2022spin}.
The pair spin magnetization is defined as the spin polarization of the condensate, ${\bm S}(\bm k)=-i{\bm d}^\ast(\bm k)\times {\bm d}(\bm k)$, integrated over the Fermi surface.
Hence, the pair spin magnetization is only possible in NUSCs.

The symmetries broken by these subsidiary orders and the observable phenomena resulting from this are summarized in Table.~\ref{subsidiaryorder}. 
Nonvanishing chirality breaks time-reversal, mirror reflection, and two-fold rotational symmetries.
These broken symmetries induce the polar-Kerr effect, the circular dichroism, the anomalous thermal Hall effect, and the anomalous acoustoelectric effect~\cite{schemm2014observation,sumiyoshi2013quantum,PhysRevB.39.8959,yip2016low,PhysRevLett.124.157002,matsushita2024impurity,yip1992circular,PhysRevB.105.134520}.
Note that these broken symmetries are necessary for the SSSE, but the superconducting state with the chirality sometimes holds the emergent two-fold spin rotational symmetry after SOC fixes the $d$-vector, where the SSSE is prohibited. 
From the broken symmetries due to the chirality, one finds the spin-chirality (spin-dependent chirality) breaks the two-fold spin rotational and momentum mirror reflection symmetries, allowing the SNE~\cite{matsushita2022spin}.
The pair spin magnetization breaks time-reversal, two-fold rotational, and two-fold spin rotational symmetries, immediately allowing the SSSE.

The pair spin polarization only exists in NUSCs~\cite{mineev2002superconducting}; however, not all such superconductors have the finite pair spin magnetization, because the average of the spin polarization of the condensate over the Fermi surface may vanish. 
Although the pair spin magnetization indeed breaks sufficient symmetries to allow the SSSE, we show below that the SSSE generally occurs in NUSCs even when the pair spin magnetization vanishes after integration over the Fermi surface.

\begin{figure}[b]
\centering
\includegraphics[width=6cm]{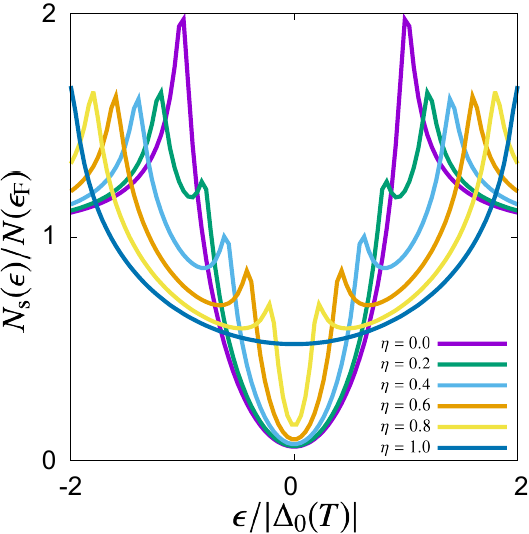}
\caption{{\bf Density of states (DOS) of the superconducting state described by Eq.~(\ref{non-uni_Weyl}).}
$N_s(\epsilon)$ is the DOS in the superconducting state and $N(\epsilon_{\rm F})$ is the DOS at the Fermi level in the normal state. 
We show the results for several values of $\eta$.
For the calculation, we set the temperature $T=0.1T_c$, the normal scattering rate $\Gamma_{\rm imp}=0.04\pi T_c$, and the normal state scattering phase shift $\delta=\pi/6$.}
\label{DOS_UCoGe}
\end{figure}

\begin{figure*}[t]
\centering
\includegraphics[width=16cm]{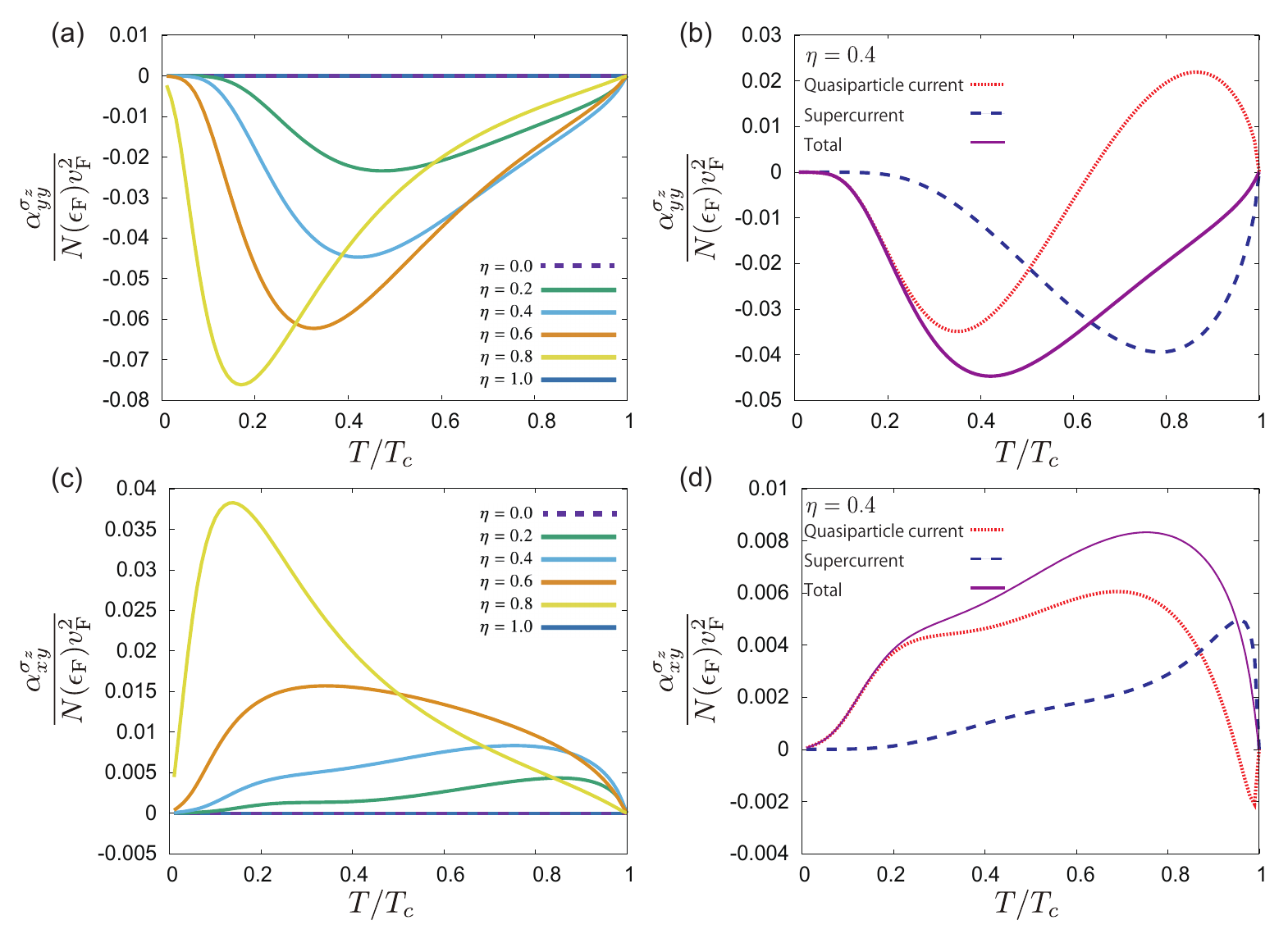}
\caption{{\bf Temperature dependence of the thermospin conductivity tensor elements of the superconducting state described by Eq.~(\ref{non-uni_Weyl}):} (a-b) spin-Seebeck conductivity, and (c-d) spin-Nernst conductivity.
In panels (a) and (c), the calculated results with several values of $\eta$ are shown.
In these panels, the curves for $\eta=0$ are shown with the dashed lines for visibility.
In panels (b) and (d), the contributions from the quasiparticle current (the red dotted curves), the supercurrent as a reaction to the dissipative electric current (the dashed blue curves), and the total of these (the solid purple curves) are shown for $\eta=0.4$. 
In all panels, we set $\Gamma_{\rm imp}=0.04\pi T_c$ and $\delta=\pi/6$ for the calculations.
}
\label{SSE_SNE_UCoGe}
\end{figure*}

\subsection*{Nonunitary superconductor with finite pair spin magnetization}
We first demonstrate that the finite pair spin magnetization, $\braket{\bm S}\neq 0$, induces the SSSE in NUSCs.
We assume a spherical Fermi surface of radius $k_{\rm F}$ and the $d$-vector of the form,
\begin{eqnarray}
\label{non-uni_Weyl}
\bm d(\bm k)=\Delta_0(T)\big(\hat{k}_x+i\hat{k}_y,\eta(\hat{k}_y-i\hat{k}_x),0\big)\, ,
\end{eqnarray}
with $\hat{\bm k}=\bm k/k_{\rm F}$. 
Without loss of generality, we take $\eta \in [0,1]$, which defines the nonunitarity and spin magnetization of Cooper pairs.
Here, $\Delta_0(T)$ is the gap amplitude, and we model its temperature dependence by 
$\sqrt{\braket{|\bm d(\bm k)|^2}_{\rm FS}}=\sqrt{2(1+\eta^2)/3}|\Delta_0(T)|=1.765 T_c \tanh(1.74\sqrt{T_c/T-1})$, where $T_c$ is the superconducting transition temperature~\cite{tinkham}.
With the $z$ axis chosen as the spin quantization axis, the superconducting order parameter matrix is diagonal in spin space and expressed as
\beq
\Delta({\bm k})=
\Delta_0(T)(\hat{k}_x+i\hat{k}_y)
\begin{pmatrix}
-1+\eta&&0\\
0&&1+\eta
\end{pmatrix}.
\eeq

We obtain the subsidiary orders generated by the superconducting order parameter in Eq.~(\ref{non-uni_Weyl}) as,
\beq
\braket{L_z}=1,\; \braket{S_zL_z}=\braket{S_z}=-\frac{2\eta}{1+\eta^2},
\eeq
and the other elements of the subsidiary orders vanish.
Equation~\eqref{non-uni_Weyl} describes the NUSCs with the pair spin magnetization when $\eta\neq 0$.
In this case, the superconducting gap amplitudes depend on the spin projection on the $z$-axis, 
$|\hat{\chi}_{\pm}(\bm k_{\rm F})|\propto |\mp 1 +\eta|$.
The splitting of the superconducting gaps is shown in Fig.~\ref{DOS_UCoGe}, where the density of states (DOS) has two distinct peaks for finite $\eta$.  
These spin-dependent gap amplitudes make nonvanishing spin-chirality and pair spin magnetization possible.

\begin{figure*}[t]
\centering
\includegraphics[width=16cm]{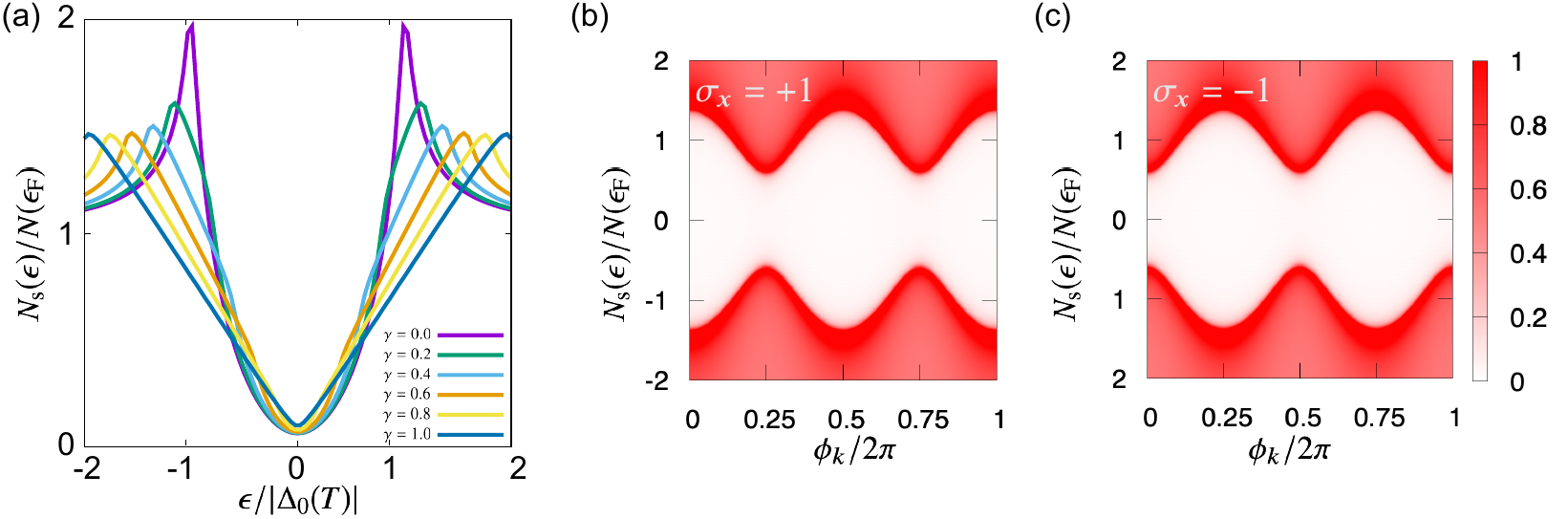}
\caption{{\bf DOS and the spin-resolved spectra of the superconducting state described by Eq.~(\ref{non-uni_Dirac}):} (a) DOS, and (b-c) spin-resolved spectra in the $k_x=0$ plane.
$N_s(\epsilon)$ is the DOS in the superconducting state and $N(\epsilon_{\rm F})$ is the DOS at the Fermi level in the normal state. 
In panel (a), we show the results for several values of $\gamma$.
In panels (b-c), we set $\gamma=0.4$ and plot the spin-resolved spectra as a function of $\phi_k=\tan^{-1}(k_y/k_z)$.
In all panels, we set $T=0.1T_c$, $\Gamma_{\rm imp}=0.04\pi T_c$ and $\delta=\pi/6$ for the calculations.
The color bar is shared between panels (b-c).
}
\label{UTe2_DOS}
\end{figure*}

The details of our linear response calculation are given in the Methods section.
Using the quasiclassical (Eilenberger) approach, we compute the dissipative electric current, $\bm J_{\rm QP}$, carried by quasiparticles driven by thermal gradients.
In the Eilenberger theory, such thermoelectric current appears via the effective particle-hole asymmetry due to the impurity potential~\cite{matsushita2022spin}, which arises from the impurity vertex corrections upon inclusion of the multiple scattering processes.
To account for the exclusion of the magnetic flux from the bulk, i.e., the vanishing of the thermoelectric charge current~\cite{ginzburg1978thermoelectric,Shelly2016},
we introduce the phase gradient into the equilibrium superconducting order according to $\Delta({\bm k}_{\rm F})\to e^{i\bm Q \cdot \bm x} \Delta({\bm k}_{\rm F})$ and choose $\bm Q$ such that ${\bm J}_{\rm QP}+{\bm J}_{\rm SC}=0$. Under this condition, the nontrivial spin response survives.

The nonunitary superconducting order given by Eq.~\eqref{non-uni_Weyl} maintains the two-fold rotational symmetry around the $z$ axis ($\mathcal{C}_2^z$). 
As discussed before, this symmetry requires $\alpha_{\mu \mu}^{\sigma_x}=\alpha_{\mu \mu}^{\sigma_y}=0$, whereas $\alpha_{\mu \mu}^{\sigma_z}$ can be finite. Figure~\ref{SSE_SNE_UCoGe} (a) demonstrates that the SSSE with the spin polarization along the $z$ axis is induced by the pair spin magnetization at finite $\eta$ (see the supplemental information for $\alpha_{xx}^z$ and $\alpha_{zz}^z$). 
Note that the unitary case $(\eta=0)$ is an example where the chirality $\left(\langle L_z\rangle=1\right)$ is not compatible with the two-fold rotational symmetry about the $x$ axis ($\mathcal{C}_2^x$), but the order parameter, Eq.~\eqref{non-uni_Weyl}, preserves the two-fold spin rotational symmetry about the $x$ axis ($\mathcal{C}_2^{\sigma_x}$), resulting in the vanishing of the corresponding transport coefficients, $\alpha_{\mu \nu}^{\sigma_z}=0$.

Note that the SSSE is also absent in the fully polarized case ($\eta =1$), where only a single gap in one spin channel remains, $\hat{\chi}_{\sigma_z=+1}=0$.
In this situation, the electric current is proportional to the spin current and the cancellation of the thermoelectric response by the supercurrents leads to vanishing of the spin current as well.
Thus, the SSSE in NUSCs exists only in cases of the partial spin polarization of the condensate. 
As an example, in Fig.~\ref{SSE_SNE_UCoGe} (b), we show that the reaction of the supercurrent does not fully cancel the spin current carried by quasiparticles when $\eta=0.4$.

As shown in Fig.~\ref{SSE_SNE_UCoGe} (b), the longitudinal quasiparticle spin current reverses direction depending on the temperature. 
This is a consequence of the different gap amplitudes for the different spin projections, $\Delta_\pm=|(1\mp\eta)\Delta_0(T)|$. 
At the low temperature, for $T \ll T_c$, the quasiparticle excitations with $\sigma_z=-1$ get suppressed by the large superconducting gap while those with $\sigma_z=+1$ dominate the SSSE response.
On the other hand, for $T \lesssim T_c$, there is an underlying competition between (1) the increase in the pairing amplitude (or gap function) and the quasiparticle lifetime (or the decrease in the imaginary part of the impurity self-energy in our formulation), and (2) the reduction in the DOS due to the superconducting gap.
In the limit of $T\gg |\Delta_0(T)|$, the latter effect plays a relatively minor role and the increase in the pairing amplitude and the quasiparticle lifetime makes the contribution of the quasiparticles with $\sigma_z=-1$ dominant.

Figure~\ref{SSE_SNE_UCoGe} (a) shows that the maximal amplitude of $\alpha_{y y}^{\sigma_z}$ shifts to lower temperatures as $\eta$ increases towards unity. This maximum is reached roughly when the temperature becomes comparable to the smaller energy gap.
At this temperature, there is a large imbalance in the quasiparticle population with the different spin projections because, under this condition, the $\sigma_z=-1$ quasiparticles are gapped while the $\sigma_z=+1$ quasiparticles are effectively gapless.

The two-fold rotational symmetry along the $z$ axis ($\mathcal{C}_2^z$) restricts the SNCs. As shown in the supplementary information, the only nonvanishing SNCs in this model are $\alpha^{\sigma_z}_{xy}=-\alpha^{\sigma_z}_{yz}$.
Figure~\ref{SSE_SNE_UCoGe} (c) demonstrates that the spin-chirality generated by the finite $\eta$ induces the SNE in NUSCs, reflecting the breaking of two-fold spin rotational and momentum mirror reflection symmetries by $\langle S_zL_z\rangle$.
Quite generally, the anomalous transverse transport arises from the asymmetry of the skew scattering of quasiparticles at impurity sites.
Such asymmetry emerges from the impurity vertex corrections when the condensate has the chirality~\cite{PhysRevB.39.8959,yip2016low, PhysRevLett.124.157002,matsushita2024impurity}, and hence the spin-chirality leads to transverse spin transport, i.e. the SNE~\cite{matsushita2022spin}. 
As shown in Fig.~\ref{SSE_SNE_UCoGe} (c), the SNE is absent when $\eta=0$ because the two-fold spin rotational symmetry about the $x$ axis ($\mathcal{C}_2^{\sigma_x}$) holds.
Note also that the anomalous Nernst effect, the generation of an electric current perpendicular to thermal gradients, is once again subject to the reaction of the supercurrent and vanishes in the bulk.
As shown in Fig.~\ref{SSE_SNE_UCoGe} (d), this reaction of the supercurrent in the anomalous Nernst effect modifies the SNE in NUSCs.
As in the SSSE, the SNE is also fully canceled by the supercurrent when the spin polarization is complete.

\begin{figure*}[t]
\centering
\includegraphics[width=16cm]{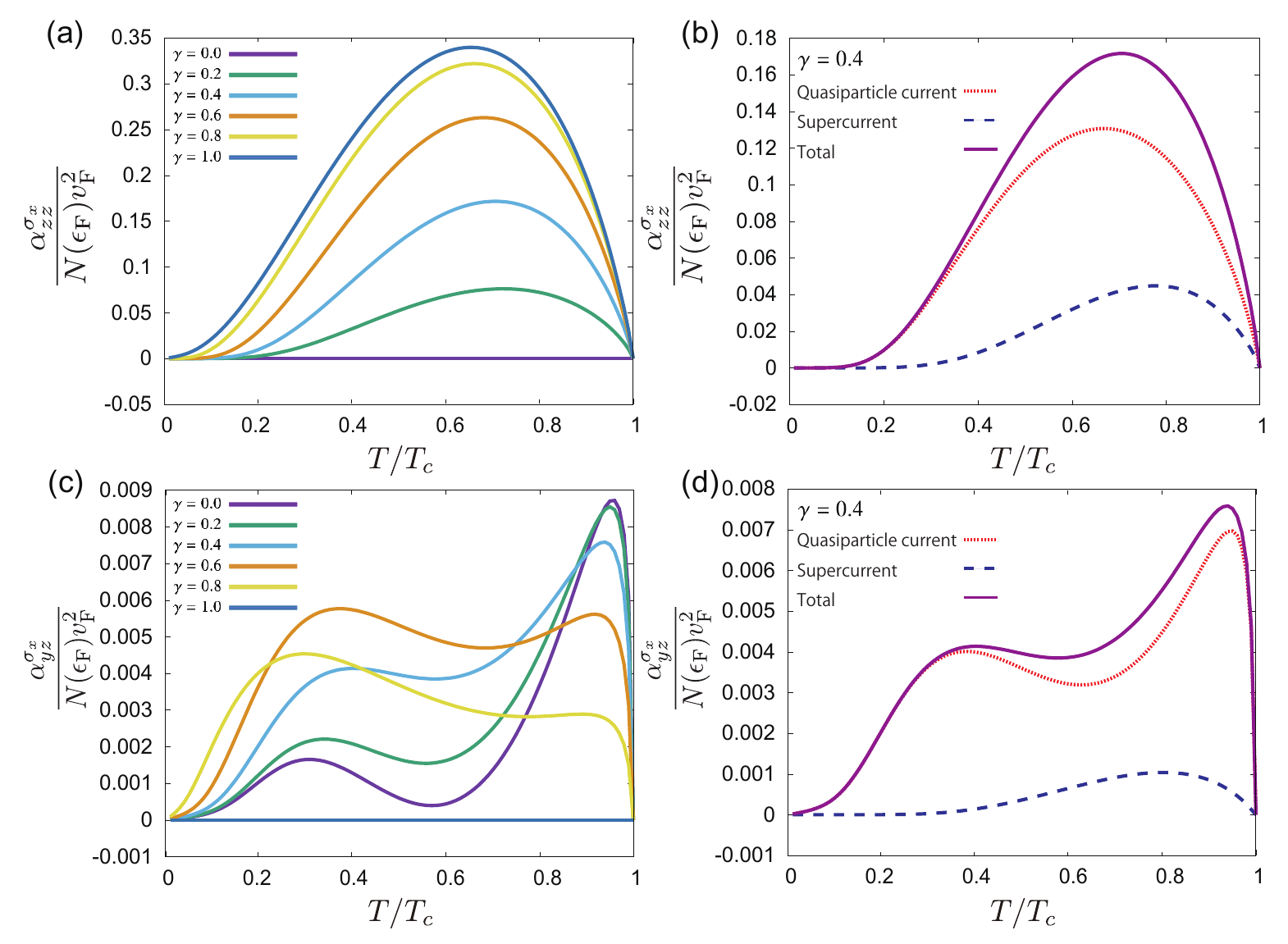}
\caption{{\bf Temperature dependence of the thermospin conductivity tensor elements of the superconducting state described by Eq.~(\ref{non-uni_Dirac}):} (a-b) the spin-Seebeck conductivity, and (c-d) the spin-Nernst conductivity.
In the panels (a) and (c), the calculated results with several values of $\gamma$ are shown.
In the panels (b) and (d), the contributions from the quasiparticle current (the red dotted curves), the supercurrent as a reaction of the dissipative electric current (the dashed blue curves), and the total of these (the solid purple curves) are shown for $\gamma=0.4$. 
In all panels, we set $\Gamma_{\rm imp}=0.04\pi T_c$ and $\delta=\pi/6$ for the calculations.}
\label{SSE_SNE_UTe2}
\end{figure*}

\subsection*{Nonunitary superconductor without the pair spin magnetization}
We finally demonstrate that the SSSE is generic in NUSCs even in the absence of the net pair spin magnetization.
As an example, we consider the following $d$-vector on the Fermi sphere,
\begin{eqnarray}
\label{non-uni_Dirac}
\bm d(\bm k)=\Delta_0(T)\big( 0,\hat{k}_z+i\gamma \hat{k}_y,\hat{k}_y+i\gamma \hat{k}_z \big),
\end{eqnarray}
and again assume $\sqrt{\braket{|\bm d(\bm k)|^2}_{\rm FS}}=\sqrt{2(1+\gamma^2)/3}|\Delta_0(T)|=1.765 T_c \tanh(1.74\sqrt{T_c/T-1})$~\cite{tinkham}. We also introduce $\gamma \in [0,1]$ as a parameter that characterizes the nonunitary Cooper pairs with a spin texture in momentum space.
This order parameter leads to different quasiparticle spectra for the spins parallel or antiparallel to the $x$ axis. Indeed, when the $x$ axis is chosen as the spin quantization axis, the superconducting order parameter matrix becomes,
\begin{align}
&\Delta(\hat{\bm k})=
\Delta_0(T)\nonumber\\
&\times
\begin{pmatrix}
(1-\gamma)\hat{k}_y+i(1+\gamma)\hat{k}_z&&0\\
0&&-(1+\gamma)\hat{k}_y+i(1-\gamma)\hat{k}_z
\end{pmatrix}.
\end{align}

When $\gamma\neq 0$, Eq.~\eqref{non-uni_Dirac} describes the nonunitary state, but the spin polarization of the Cooper pairs,
$S_x(\bm k)\propto (\hat{k}_z^2-\hat{k}_y^2)$, vanishes after the integration over the Fermi surface.
As shown in Fig.~\ref{UTe2_DOS}, the coherence peaks in the DOS do not split (see panel (a)), but the spectrum depends on the spin (see panels (b-c)).
For $\gamma>0$, the $\sigma_x=+1\;(\sigma_x=-1)$ spin pairing sector has the low energy excitations along the $y\;(z)$ direction.

The nonunitary superconducting order described by Eq.~\eqref{non-uni_Dirac} respects the two-fold rotational symmetry along the $x$ axis ($\mathcal{C}_2^x$), which requires $\alpha_{\mu \mu}^{\sigma_y} = \alpha_{\mu \mu}^{\sigma_z} = 0$, whereas $\alpha_{\mu \mu}^{\sigma_x}$ can be finite. Figure~\ref{SSE_SNE_UTe2} (a) demonstrates that this spin-dependent spectrum induces the SSSE with the spin polarization along the $x$ axis.
The thermal gradient along the $z$ direction excites quasiparticles in the $\sigma_x= -1$ sector more than those in the $\sigma_x= +1$ sector, giving rise to the SSSE after taking into account the cancellation of the electric current. 
This illustrates the essential point of our analysis: in NUSCs, the quasiparticle spectrum along a given direction is different for opposite spin projections on the spin quantization axis. 
Hence, when we compute the transport coefficients weighted by the projection of the group velocity along or perpendicular to the thermal gradient, the contributions of the two spin sectors do not compensate each other.
In general, the spin and electric currents differ for a given direction of the thermal gradient, so that the SSSE emerges even when the thermoelectric current vanishes due to the reaction of the supercurrent. 
Note that the supercurrent still carries spin even in the absence of net spin magnetization. 
This behavior is shown in Fig.~\ref{SSE_SNE_UTe2} (b), where the reaction of the supercurrent modifies the SSSE even when pair spin magnetization is absent.

Figure~\ref{SSE_SNE_UTe2} (a) shows that the SSSE is enhanced as $\gamma\to 1$.
As shown in Fig.~\ref{SSE_SNE_UTe2} (b), the quasiparticle spin current dominates the SSSE.
Hence, this amplification of the SSSE should be understood from the quasiparticle spectrum.
As shown in Fig.~\ref{UTe2_DOS} (b-c), the spectra have the gap minima in the $y$ or $z$ direction. These excitation gaps completely vanish at $\gamma=1$, forming line nodes.
The formation of line nodes significantly enhances quasiparticle excitations, resulting in the large SSC as $\gamma\to 1$.

\begin{figure}[t]
\centering
\includegraphics[width=8cm]{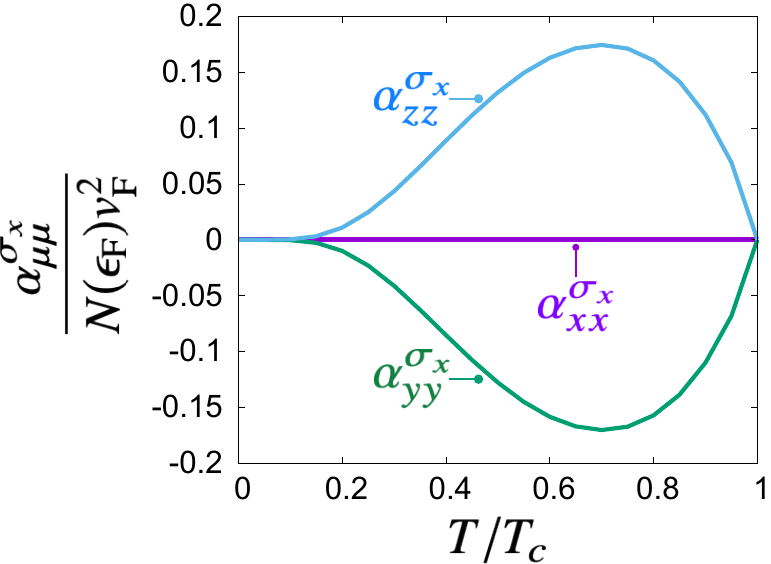}
\caption{
{\bf Temperature dependence of all SSCs not prohibited by the crystalline symmetries in the superconducting state described by Eq.~(\ref{non-uni_Dirac}).}
For the calculation, we set $\gamma=0.4$, $\Gamma_{\rm imp}=0.04\pi T_c$ and $\delta=\pi/6$ for the calculations.
}
\label{tempdep_axmm}
\end{figure}

In contrast to the nonunitary state described by Eq.~\eqref{non-uni_Weyl}, the quasiparticle contribution does not change the sign of the thermospin conductivity tensor below $T_c$ because the gap amplitude does not depend on the spin projection.
However, as Fig.~\ref{UTe2_DOS} (b-c) show, for a given direction in the $yz$ plane, the DOS is dominated by one of the spin projections.
Therefore, the direction of the thermal gradient determines the spin orientation yielding a greater contribution.
Indeed, as shown in Fig.~\ref{tempdep_axmm}, the sign of the SSCs depends on the direction of the thermal gradient in the $yz$ plane, and $\alpha_{zz}^x=-\alpha_{yy}^x$ is satisfied.
This result is consistent with the spin polarization in momentum space given by $S_x(\bm k)\propto (\hat{k}_z^2-\hat{k}_y^2)$.
Figure~\ref{tempdep_axmm} also shows $\alpha_{xx}^{\sigma_x}=0$, reflecting that the thermal gradient along the $x$ direction preserves the population balance of quasiparticles between the opposite spin sectors.

The two-fold rotational symmetry along the $x$ axis ($\mathcal{C}_2^x$) restricts the nonvanishing SNCs only to $\alpha^{\sigma_x}_{yz}=-\alpha^{\sigma_x}_{zy}$, as shown in Fig.~\ref{SSE_SNE_UTe2} (c-d).

\subsection*{Experimental detection}

\begin{figure*}[t]
\centering
\includegraphics[width=18cm]{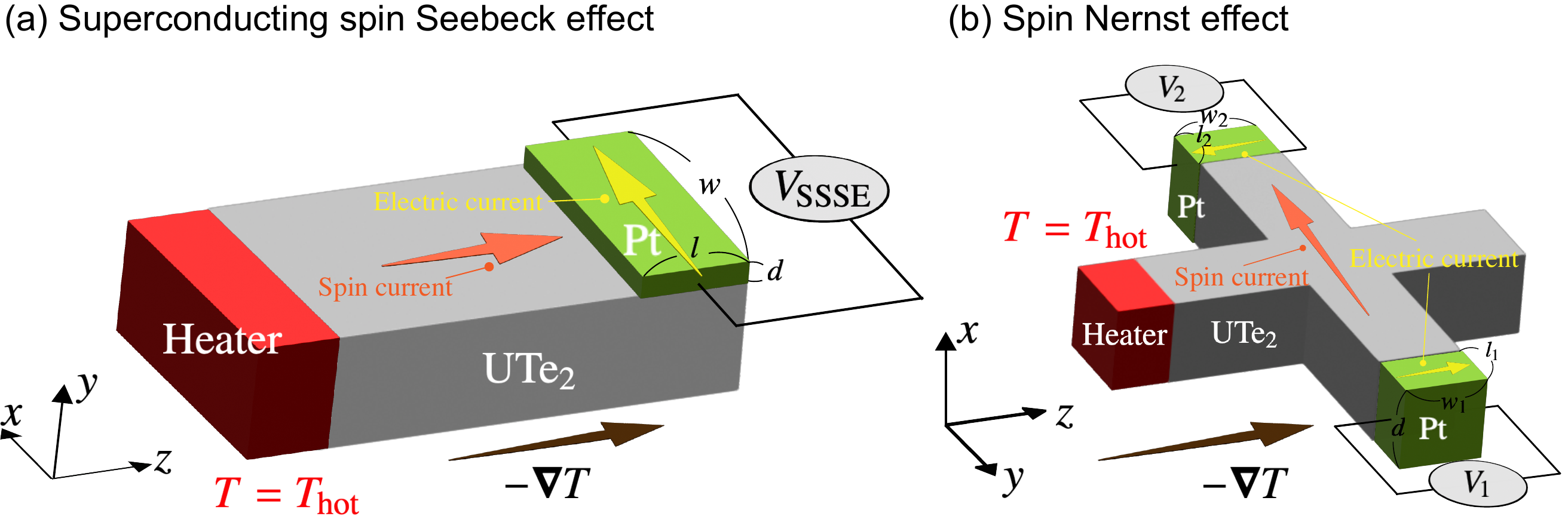}
\caption{
{\bf Sketches of experimental setups to detect the superconducting spin caloritronics phenomena in NUSCs :} (a) the SSSE and (b) the SNE. The sample of NUSCs (such as UTe$_2$, shown in grey) is connected to a detector of spin current with a large spin Hall angle (such as Pt, shown in green). The orange (yellow) arrows represent the spin (electric) current induced by the SSSE (ISHE), and the brown arrows represent thermal gradients along the $z$-direction, respectively.
}
\label{setup}
\end{figure*}

Having demonstrated the existence of the SSSE and the SNE in NUSCs, we now examine the feasibility of detecting their signals.
The magnitude of the spin current density is often estimated as the electric current density by multiplying by $e/\hbar$. 
As an example, we consider the nonunitary B$_{3u}+i$A$_u$ state, which is a possible candidate for the superconducting order in UTe$_2$~\cite{ishihara2023chiral}, and assume the applied thermal gradient $\Delta T=0.1$ K in $1.0$ mm along the $z$-direction. We take the following parameters for UTe$_2$: the Fermi velocity $v_{\rm F}=|\bm v_{\rm F}|=1.5\times 10^4$ m/s, the superconducting transition temperature $T_c=2.0$ K, and the normal state DOS at the Fermi energy $N(\epsilon_{\rm F})=1.2\times 10$ eV$^{-1}$ ~\cite{aoki2022unconventional}.
With these values, the magnitudes of the spin currents at $T=1.6$ K are estimated as $7.1 \times 10^2$ $\mu$A/cm$^2$ for the SSSE and $4.7 \times 10$ $\mu$A/cm$^2$ for the SNE.

The spin current is not directly measurable but is usually detected by converting it into electric voltage through the inverse spin Hall effect (ISHE)~\cite{sinova2015spin}.
Typical setups for detecting the spin current are schematically shown in Fig.~\ref{setup}. 
These setups are similar to those for the spin Seebeck effect in magnonic systems or the spin Nernst effect in heavy metals~\cite{uchida2008observation,bose2018,sheng2017}.
The spin current generated by the SSSE or the SNE in UTe$_{2}$, a candidate of NUSCs, is absorbed into the detector of the spin current and converted into the electric current via the ISHE, resulting in the electric voltage ($V_{\rm SSSE}$ in panel (a) and $V_1$ and $V_2$ in panel (b)). 
We consider Pt as a detector because it is widely used in the field of spintronics.
Pt has a spin Hall angle $\theta_{\rm SHA}=4.4\times 10^{-3}$ and the resistivity $\rho=2.0\times 10$ n$\Omega\cdot$m~\cite{PhysRevLett.99.226604,sinova2015spin}.
Our qualitative estimate of the voltage signal for the SSSE is $V_{\rm SSSE}\simeq 1.9 \times 10$ nV, for $w=1.0 \times 10^2$ $\mu$m and $l=1.0 \times 10$ $\mu$m, and $d=1.0 \times 10$ nm in Fig.~\ref{setup} (a).
Similarly, we estimate  $V_{\rm SNE}\equiv V_1-V_2\simeq 2.1$ nV as the signal of the SNE when we take $w_1=w_2=1.0 \times 10^2$ $\mu$m, $l_1=l_2=1.0 \times 10^2$ nm, and $d=2.0$ $\mu$m  in Fig.~\ref{setup} (b).
These estimated voltage signals are large enough to be measured, and compare favorably with the signals that have been experimentally detected in other spin heterostructures. Hence we believe that the experimental detection of the SSSE and the SNE in NUSCs is feasible.

We note that the above estimation does not take into account the scattering events at the interface between the NUSC candidate and the detector.
The spin current generated within NUSCs is subjected to the Andreev scattering at the interface.
This scattering event involves the conversion of quasiparticles into Cooper pairs and vice versa, and includes the rotation of the spin polarization, which may modify the spin current injected into the detector from NUSCs.
The evaluation of these is beyond the scope of 
the current proof-of principle treatment, and remains a subject for future study.

The challenges in detecting the SSSE and the SNE include fabricating microdevices with the NUSC candidate materials (most of which are uranium compounds). Recently, magnetoresistivity was measured in microdevices fabricated with UTe$_2$~\cite{helm2024field}, supporting our argument that the measurements, so that we propose are achievable in the near future.

\section*{Discussion}
The order parameters considered above are motivated by recent proposals relevant to the intensely studied unconventional superconductors.
The $d$-vector in Eq.~(\ref{non-uni_Weyl}) is established in the A$_1$ and A$_2$ phases of the superfluid $^3$He under applied magnetic fields~\cite{vollhardt2013superfluid}. In the superfluid $^3$He, the interplay between unconventional pairing and quasiparticle impurity scattering can be engineered by high-porosity silica aerogel~\cite{halperin}. The aerogel, which is modeled by randomly distributed nonmagnetic impurities with $\delta \approx \pi /2$ and $\Gamma_{\rm imp}\approx 0.1$ -- $0.2\pi T_{\rm c}$~\cite{thun98}, realizes the A$_1$ and A$_2$ phases under magnetic fields~\cite{choi04}. Although the calculated results with Eq.~(\ref{non-uni_Weyl}) can be applied to these phases, the effects of the spin supercurrent and the collective motion of the $d$-vector have yet to be explored.

The $d$-vector in Eq.~(\ref{non-uni_Weyl}) has been suggested for UCoGe, where superconductivity
coexists with ferromagnetically ordered moments along the $z$ direction~\cite{aoki2019review}.
Because superconductivity emerges from this ferromagnetic state, the spin of the condensate must be aligned along the same direction, so that $\braket{S_z}\neq 0$ is realized~\cite{mineev2002superconducting}.
The NMR measurement supports the realization of the A$_1$ state described by $\bm d(\bm k)= (a_1k_x+ia_2k_y,a_3k_y+ia_4k_x,0)$, where $a_i\;(i=1,2,3,4)$ are real coefficients~\cite{hattori2012superconductivity}.
Equation~(\ref{non-uni_Weyl}) is obtained if we set $a_1=a_2,\; a_3/a_1=-a_4/a_1=\eta$ and thus our results with Eq.~(\ref{non-uni_Weyl}) can be directly applied to UCoGe.

The main caveat is that ferromagnetic superconductors may support a spontaneous vortex state~\cite{ohta2010microscopic,kusunose2013theory}.
In such cases, the Meissner screening is not only incomplete, but vortices generate unpaired quasiparticles, and vortex motion contributes to the transverse transport. Nonetheless, in the majority of ferromagnetic superconductors, the evidence for the spontaneous vortex state is scant, and thus we believe that our results are applicable. 
If the exchange splitting is large enough to make the difference in e.g. the Fermi velocities between the two spin orientations substantial, we may expect corrections to the quasiclassical results. However, in most materials under discussion, the splitting is small and appears as a higher-order correction in the powers of $(k_{\rm F}\xi_0)^{-1}\ll 1$, where $\xi_0$ is the superconducting coherence length.
In ferromagnetic superconductors, in principle, magnons also contribute to the spin transport induced by thermal gradients.
The magnitude of this contribution depends on the magnon spectrum. 
In superconductors such as UCoGe and URhGe, the exchange interaction is Ising-like,~\cite{hattori2012superconductivity,aoki2019review}, and hence the magnons are gapped. 
Because the superconducting transition temperature is far lower than the magnon gap, the corresponding contribution is exponentially suppressed at the temperatures of interest.

Another notable candidate for spin-triplet superconductivity is UTe$_2$, which exhibits a very large upper critical magnetic field (over $30$ T) and strong magnetic Ising anisotropy~\cite{aoki2022unconventional,ran2019nearly,knebel2019field}. These features strongly support spin-triplet superconductivity in UTe$_2$. Early on there were indications that nonunitary superconductivity may be realized in this compound. Observation of polar Kerr signal below $T_c$ suggests realization of a superconducting order with the broken time-reversal symmetry~\cite{hayes2021multicomponent}. Furthermore, the magnetic field penetration depth measurement supports the realization of the B$_{3u}+i$A$_u$ state, which is nonunitary~\cite{ishihara2023chiral}. However, more recent experiments with high-quality samples failed to observe the polar Kerr signal in UTe$_2$ in the absence of magnetic fields, and therefore currently the nonunitary superconductivity in UTe$_2$ is still a subject of debate~\cite{PhysRevX.13.041019}. Our results above may help to settle this question.
Note that the B$_{3u}+i$A$_u$ state is described by $\bm d(\bm k)= (i a_1k_x,b_1k_z+ia_2k_y,b_2k_y+ia_3k_z)$, where $a_i\;(i=1,2,3)$ are the coefficients for the basis function of the A$_u$ representation and $b_i\;(i=1,2)$ are coefficients for the basis function of the B$_{3u}$ representation, respectively~\cite{ishihara2023chiral}. Recent theoretical analysis argued that strong magnetic Ising anisotropy suppresses the $x$ component of the $d$-vector in UTe$_2$~\cite{tei2023pairing}. When we combine these results and then set $a_1=0,\;b_2=b_3,\;a_2/b_2=a_3/b_3=\gamma$, the $d$-vector reduces to Eq.~(\ref{non-uni_Dirac}). Consequently, our model is directly applicable to the candidate orders in UTe$_2$.
Note that UTe$_2$ is paramagnetic above $T_c$ (albeit it may be close to a ferromagnetic phase), which excludes the spontaneous vortex state. The SSSE is thus useful as a test for nonunitary superconductivity in UTe$_2$.

As shown in Table~\ref{subsidiaryorder}, most NUSC candidates are uranium compounds, where SOC is strong. SOC fixes the direction of the $d$-vector and determines the specific form of the superconducting order parameter, which is therefore accounted for in our calculations. 
At the same time, anti-symmetric SOC, which takes a staggered form to maintain global inversion symmetry, arises due to the lack of an inversion center at each U-site~\cite{aoki2019review,aoki2022unconventional}. This anti-symmetric SOC hybridizes microscopic degrees of freedom and introduces additional contributions to the SSSE and the SNE~\cite{fujimoto2024microscopic,matsushita2024intrinsic}.
These additional contributions from the anti-symmetric SOC appear already above the superconducting transition temperature and become most prominent when $T\sim E_{\rm  SOC}$, where $E_{\rm  SOC}$ is the magnitude of the anti-symmetric SOC and typically $E_{\rm SOC}=\mathcal{O}(10^3{\rm K})\gg T_c$ in uranium-based superconductors.
Because the contributions from the anti-symmetric SOC are peaked at $T\sim E_{\rm  SOC}$ and get suppressed below this temperature, the superconducting response dominates the SSSE and the SNE at $T\lesssim T_c \ll E_{\rm  SOC}$. 

Both in the normal and superconducting states, the SNCs have two contributions.
One is the spin current carried by quasiparticles (or normal electrons above $T_c$) in response to the thermal gradient~\cite{matsushita2022spin}.
The second contribution depends on whether the material is in the normal or superconducting state.
In the normal state, the Seebeck voltage induces the spin Hall effect and modifies the SNE~\cite{fujimoto2024microscopic,matsushita2024intrinsic}, whereas in the superconducting state, the supercurrent compensates the thermoelectric current, changing the SNE.

In our calculations, the existence of the SSSE and the SNE relies on the asymmetries of impurity scattering due to the structure of the order parameter. 
In principle, the particle-hole asymmetry in the normal state leads to similar asymmetries~\cite{yip1992circular,PhysRevB.105.134520} and may dominate for sufficiently clean samples. 
However, in the weak coupling regime, this effect is again of order $(k_{\rm F}\xi_0)^{-1}\ll 1$, and therefore in realistic materials, the impurity mechanism is clearly dominant. 

Our conclusions remain qualitatively valid for multiband and multiorbital systems, but at a quantitative level, the SSSE and the SNE depend on the specific structure of the energy bands and other material details. 
In principle, in such systems, an impurity-independent mechanism for the SNE is allowed due to a geometric quantity called the spin Berry curvature~\cite{fujimoto2024microscopic,matsushita2024intrinsic}. 
This mechanism reflects the geometric nature of momentum space and is beyond the quasiclassical approximation. 
We expect that in many candidate NUSC materials, the impurity contribution is dominant, but the relative magnitude of the two effects remains to be explored in future studies.

In summary, we showed that the SSSE is a generic feature of NUSCs, irrespective of whether the superconducting condensate has net pair spin magnetization. We placed our results in the context of symmetries broken by the subsidiary classification of the order parameters with respect to the chirality, the spin-chirality, and the pair spin magnetization. 
Together with the SNE, which is transverse counterpart of the SSSE, and sensitive to the spin-chirality and tests for the helical superconducting order in time-reversal-invariant topological superconductors~\cite{matsushita2022spin}, our work establishes spin caloritronic phenomena as sensitive probes of the symmetries of the order parameters in spin-triplet superconductors.
We argued for the possible relevance of our detailed calculations to measurements in UTe$_2$ and uranium-based ferromagnetic superconductors, but our broader conclusions remain applicable to a wide range of currently known and future NUSCs.

\section*{Materials and Methods}
\subsection*{Quasiclassical Eilenberger theory of superconductivity}
The quasiclassical method in the theory of superconductivity relies on the small parameter, $(k_{\rm F}\xi_0)^{-1}\sim T_c/\epsilon_{\rm F}$, in superconductors, where $\epsilon_{\rm F}$ is the Fermi energy~\cite{eilenberger1968transformation,serene1983quasiclassical}, to develop an approximation scheme for the Green function. 
To formulate the spin current responses to the thermal gradient, we consider the Green function, $\check{G}$, in the spin, particle-hole (Nambu), and time-ordered Keldysh space. 
When $(k_{\rm F}\xi_0)^{-1}\ll 1$, all the elements of the Green function are sharply peaked at the Fermi energy and weakly depend on the energy far away from it.  
The high-energy (away from the Fermi surface) part of $\check{G}$ renormalizes interactions between electrons. 
The low-energy part determines the equilibrium properties and response to external perturbations. The essence of this method is to define the Keldysh quasiclassical Green function (QGF), $\check{g}(\epsilon,{\bm k}_{\rm F})$, as the integral of $\check{G}$ over energy, so that the major contribution comes from the vicinity of the Fermi surface.

To implement this approach, we linearize the dispersion of normal electrons near the Fermi energy, $\xi_{\bm k}={\bm v}_{\rm F}\cdot({\bm k}-{\bm k}_{\rm F})$, 
with the Fermi velocity ${\bm v}_{\rm F}\equiv (\partial\xi_{\bm k}/\partial {\bm k})_{{\bm k}={\bm k}_{\rm F}}$, and then the QGF, $\check{g}$, is defined in the Keldysh space as~\cite{serene1983quasiclassical}, 
\begin{align}
\check{g}(\epsilon,{\bm k}_{\rm F})
= 
\begin{pmatrix}
\underline{g}^{{\rm R}}(\epsilon,{\bm k}_{\rm F})&\underline{g}^{{\rm K}}(\epsilon,{\bm k}_{\rm F})\\
0& \underline{g}^{{\rm A}}(\epsilon,{\bm k}_{\rm F})
\end{pmatrix}
\equiv
\int_{-\epsilon_{\rm c}}^{\epsilon_{\rm c}} d\xi_{\bm k}  \check{\tau}_z\check{G}(\epsilon,{\bm k}),
\label{Greenfunc_Kspace}
\end{align}
where $\check{\tau}_\mu\;(\mu=x,y,z)$ are the Pauli matrices in the particle-hole space and $\epsilon_c$ is the cutoff energy satisfying $T_c \ll \epsilon_c \ll \epsilon_{\rm F}$.
The superscripts, ${\rm X}={\rm R}, {\rm A}, {\rm K}$, represent the retarded, advanced and Keldysh components.
Each of these components, in turn, is a matrix in the spin and particle-hole space,
\begin{align}
\label{Greenfunc_Nspace}
\underline{g}^{\rm X}
=
\begin{pmatrix}
g^{\rm X} + {\bm g}^{\rm X}\cdot{\bm \sigma}& 
[\bm{\sigma} \cdot {\bm f}^{\rm X}](i\sigma_y)\\
(i \sigma_y)[\bm \sigma \cdot \bar{{\bm f}}^{\rm X}]
& \bar{g}^{{\rm X}} - \sigma_y \bar{\bm g}^{\rm X}\cdot{\bm \sigma}\sigma_y
\end{pmatrix},
\end{align}
where ${\bm \sigma}=(\sigma_x,\sigma_y,\sigma_z)$ is the vector of the Pauli matrices in spin space.
$g^{\rm X}$ and $\overline{g}^{{\rm X}}$ (${\bm g}^{{\rm X}}$ and $\bar{\bm g}^{{\rm X}}$) are the spin scalar (vector) components of the electron and hole propagators, ${\bm f}^{\rm X}$ and $\bar{{\bm f}}^{\rm X}$ are the spin-triplet electron and hole pair amplitudes, respectively. 
Throughout the manuscript, we denote a matrix in the Keldysh space as $\check{A}$ and a matrix in the spin and particle-hole space as $\underline{A}$.

The QGF obeys the standard Eilenberger equation,
\begin{align}
\left[\epsilon \check{\tau}_z- \check{\Delta}({\bm k}_{\rm F})-\check{\sigma}_{\rm imp}, \check{g}\right]+i{\bm v}_{\rm F}\cdot {\bm \nabla} \check{g}=0\,,
\label{eq:transport_hom}
\end{align}
and is supplemented by the normalization condition~\cite{eilenberger1968transformation,serene1983quasiclassical},
\beq
\check{g}^2=-\pi^2.
\label{eq:normalization}
\eeq
This normalization condition is derived from the fact that the square of the QGF also satisfies Eq.~\eqref{eq:transport_hom}~\cite{kopnin2001theory}. This condition must be satisfied irrespective of the structure of the order parameter~\cite{larkin1969quasiclassical,uematsu}.
$\check{\Delta}({\bm k}_{\rm F})$ in Eq.~\eqref{eq:transport_hom} is the mean-field pairing self-energy.
In spin-triplet superconductors, $\check{\Delta}({\bm k}_{\rm F})$ is associated with the $d$-vector as,
\begin{subequations}
\beq
\check{\Delta}({\bm k}_{\rm F})&=&
\begin{pmatrix}
\underline{\Delta}({\bm k}_{\rm F})&0\\
0&\underline{\Delta}({\bm k}_{\rm F})
\end{pmatrix},\\
\underline{\Delta}({\bm k}_{\rm F})&=&
\begin{pmatrix}
0&\Delta({\bm k}_{\rm F})\\
-\Delta^\dag({\bm k}_{\rm F})&0
\end{pmatrix}\nonumber\\
&=&
\begin{pmatrix}
0&{\bm d}({\bm k}_{\rm F}) \cdot {\bm \sigma}(i\sigma_y)\\
{\bm d}^\ast({\bm k}_{\rm F}) \cdot (i\sigma_y){\bm \sigma}&0
\end{pmatrix},
\eeq
\end{subequations}
where $\Delta({\bm k}_{\rm F})={{\bm d}({{\bm k}_{\rm F})} \cdot \bm \sigma}(i\sigma_y)$ is the order parameter matrix in spin space.
When the $d$-vector satisfies $\bm d^{\ast}(\bm k) \times \bm d(\bm k)={\bm 0}$, the pairing states are referred to as unitary because $\Delta^\dag({\bm k}_{\rm F})\Delta({\bm k}_{\rm F})=|{\bm d}({\bm k}_{\rm F})|^2$.
Otherwise, the pairing states are referred to as nonunitary with $\Delta^\dag({\bm k}_{\rm F})\Delta({\bm k}_{\rm F})=|{\bm d}({\bm k}_{\rm F}))|^2+i({\bm d}^*({\bm k}_{\rm F})\times {\bm d}({\bm k}_{\rm F}))\cdot {\bm \sigma}$~\cite{sigrist1991phenomenological}.
As shown in Eq.~\eqref{spinmagnetization}, ${\bm S}(\bm k)=-i\bm d^{\ast}(\bm k) \times \bm d(\bm k)$ describes the spin polarization of the condensate in momentum space and lifts the spin-degeneracy in the quasiparticle spectrum. 

The impurity effects are incorporated in the impurity self-energy, $\check{\sigma}_{\rm imp}$.
In the quasiclassical theory, the impurity effects are essential for obtaining nonvanishing SSSE and SNE. These transport coefficients require particle-hole asymmetry. While this asymmetry may arise simply from the slope of the DOS~\cite{yip1992circular,PhysRevB.105.134520}, such effects are small for $(k_{\rm F}\xi_0)^{-1}\ll 1$, and are neglected in the Eilenberger theory. However, in superconductors, the particle-hole asymmetry also appears via the impurity vertex corrections with the inclusion of the multiple scattering processes~\cite{matsushita2022spin}.

\begin{figure}[t]
\centering
\includegraphics[width=8cm]{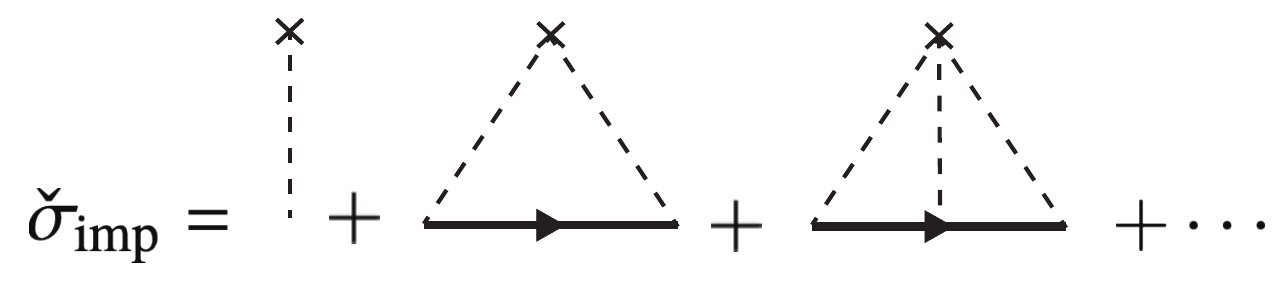}
\caption{{\bf Feynman diagrams for the impurity self-energy with the self-consistent $T$-matrix approximation.}}
\label{Tmat}
\end{figure}

We adopt the self-consistent $T$-matrix approximation (SCTMA) to compute the impurity self-energy, which contains all of the non-crossing diagrams associated with the multiple scattering processes (see Fig.~\ref{Tmat}). 
We assume randomly distributed nonmagnetic impurities with the short-range impurity potential with the potential strength, $V_{\rm imp}$, and the impurity density, $n_{\rm imp}$. 
In the SCTMA, the impurity self-energy is~\cite{balatsky2006impurity},
\begin{align}
\label{eq:tmatrix1}
\check{\sigma}_{\rm imp}=
\begin{pmatrix}
\underline{\sigma}_{\rm imp}^{{\rm R}}&\underline{\sigma}_{\rm imp}^{{\rm K}}\\
0&\underline{\sigma}_{\rm imp}^{{\rm A}}
\end{pmatrix} 
=-\Gamma_{\rm imp} \left(\cot \delta+\left\langle\frac{ \check{g}}{\pi}\right\rangle_{{\rm FS}} \right)^{-1},
\end{align}
where $\Gamma_{\rm imp}=n_{\rm imp}/(\pi N(\epsilon_{\rm F}))$ is the normal-state scattering rate and $\cot\delta = -1/(\pi N(\epsilon_{\rm F})V_{\rm imp})$ is the scattering phase shift.
These quantities are defined in the normal state and 
parameterize impurity scattering in our calculations.

We consider the dilute impurity limit, $\Gamma_{\rm imp} \ll T_{\rm c}$. 
Spin-triplet superconductivity is fragile with respect to the disorder, but for $\Gamma_{\rm imp} \ll T_{\rm c}$, the reduction of $T_{\rm c}$ is small even when the impurity potential is strong, $\cot\delta\rightarrow 0$~\cite{PhysRevB.37.83}.

\subsection*{Response to a thermal gradient}
We now formulate the response theory to the thermal gradient within the framework of the Eilenberger theory~\cite{PhysRevB.53.15147, PhysRevB.75.224502}.
The formalism is simplified in the cases when the $d$-vector has only two components as in Eq.~\eqref{non-uni_Weyl} and Eq.~\eqref{non-uni_Dirac}. In these cases, we can choose a spin quantization axis along which the order parameter matrix is diagonalized in spin space.
For example, for ${\bm d}(\bm k)=(d_x (\bm k),d_y (\bm k),0)$, the appropriate quantization axis is the $z$ axis. 
Let us denote such an axis as $\mu$ and then the QGF is also block-diagonal in spin space, $\check{g}=\check{g}^{\sigma_\mu=+1} \oplus \check{g}^{\sigma_\mu=-1}$, because both nonmagnetic impurities and the thermal gradient do not break the spin conservation.

In each spin sector, we separate $\check{g}^{\sigma_\mu}$ into the equilibrium part, $\check{g}_{\rm eq}^{\sigma_\mu}$, and its deviation from the equilibrium, $\delta \check{g}^{\sigma_\mu}=\check{g}^{\sigma_\mu}-\check{g}_{\rm eq}^{\sigma_\mu}$, which linearly depends on the thermal gradient.
The equilibrium QGF obeys the homogeneous Eilenberger equation, 
\begin{align}
\left[\epsilon \check{\tau}_z- \check{\Delta}^{\sigma_\mu}_{\rm eq}({\bm k}_{\rm F})-\check{\sigma}^{\sigma_\mu}_{\rm imp, eq}, \check{g}_{\rm eq}^{\sigma_\mu}\right]=0\,,
\label{eq:transport_eq}
\end{align} 
where the brackets denote a commutator. This equation is supplemented by the normalization condition, $[\check{g}_{\rm eq}^{{\sigma_\mu}}]^2=-\pi^2$. 
Here, $\check{\Delta}^{\sigma_\mu}_{\rm eq}$ is the mean-field pairing self-energy in the spin sector $\sigma_\mu$ and is associated with $\check{\Delta}_{\rm eq}$ as $\check{\Delta}_{\rm eq}=\check{\Delta}^{\sigma_\mu=+1}_{\rm eq} \oplus \check{\Delta}^{\sigma_\mu=-1}_{\rm eq}$. $\check{\Delta}^{\sigma_\mu}_{\rm eq}$ has the form,
\begin{subequations}
\beq
\check{\Delta}^{\sigma_\mu}_{\rm eq}({\bm k}_{\rm F})&=&
\begin{pmatrix}
\underline{\Delta}^{\sigma_\mu}_{\rm eq}({\bm k}_{\rm F})&0\\
0&\underline{\Delta}^{\sigma_\mu}_{\rm eq}({\bm k}_{\rm F})
\end{pmatrix},\\
\underline{\Delta}^{\sigma_\mu}_{\rm eq}({\bm k}_{\rm F})&=&
\begin{pmatrix}
0&\Delta^{\sigma_\mu}_{\rm eq}({\bm k}_{\rm F})\\
-\Delta^{{\sigma_\mu}\ast}_{\rm eq}({\bm k}_{\rm F})&0
\end{pmatrix}\,,
\eeq
\end{subequations}
where $\Delta^{\sigma_\mu}_{\rm eq}({\bm k}_{\rm F})$ is the equilibrium gap function in the spin sector.
In the SCTMA, the equilibrium impurity self-energy is obtained simply by using the corresponding QGF in Eq.~\eqref{eq:transport_eq}, so that in each spin sector,
\begin{align}
\label{eq:tmatrix}
\check{\sigma}^{\sigma_\mu}_{\rm imp, eq}
=-\Gamma_{\rm imp} \left[\cot \delta+\left\langle\frac{ \check{g}_{\rm eq}^{\sigma_\mu}}{\pi}\right\rangle_{{\rm FS}} \right]^{-1}.
\end{align}
The solution for the equilibrium QGF is,
\begin{subequations}
\beq
\label{eq:GRA_eq}
\underline{g}_{\rm eq}^{\sigma_\mu {\rm R,A}}&=&-\pi \frac{\underline{M}^{\sigma_\mu {\rm R,A}}}{D^{\sigma_\mu {\rm R,A}}},\\
\underline{g}_{\rm eq}^{\sigma_\mu {\rm K}}&=&(\underline{g}_{\rm eq}^{\sigma_\mu {\rm R}}-\underline{g}_{\rm eq}^{\sigma_\mu {\rm A}})\tanh(\frac{\epsilon}{2T}),
\eeq
\label{eq:g_equil}
\end{subequations}
where the numerator, $\underline{M}^{\sigma_\mu {\rm R, A}}=\tilde{\epsilon}^{\sigma_\mu {\rm R, A}}\underline{\tau}_z-\underline{\Delta}_{\rm eq}^{\sigma_\mu}$, and the denominator, $D^{\sigma_\mu {\rm R, A}}=\sqrt{|\Delta^{\sigma_\mu}_{\rm eq}({\bm k}_{\rm F})|^2-\tilde{\epsilon}^{\sigma_\mu {\rm R, A}\;2}}$, are defined with the renormalized energy,  $\tilde{\epsilon}^{\sigma_\mu {\rm R, A}}=\epsilon-\frac{1}{2}{\rm Tr}(\underline{\tau}_z \underline{\sigma}_{\rm imp, eq}^{\sigma_\mu {\rm R, A}})$.
Note that $\check{\sigma}^{\sigma_\mu}_{\rm imp, eq}$ is diagonal in the particle-hole space because $\braket{\Delta^{\sigma_\mu}_{\rm eq}({\bm k}_{\rm F})}_{\rm FS}=0$ is satisfied in spin-triplet superconductors.

The equilibrium QGF obtained above is spatially uniform. 
If the thermal gradient driving the system away from equilibrium results in the slow spatial dependence of the quasiparticle population, we can develop a perturbation theory for the thermal gradient. In this case, the nonequilibrium correction to the QGF obeys the equation,
\begin{align}
&\left[\epsilon \check{\tau}_z- \check{\Delta}^{\sigma_\mu}_{\rm eq}({\bm k}_{\rm F})-\check{\sigma}^{\sigma_\mu}_{\rm imp, eq}, \delta \check{g}^{\sigma_\mu }\right]\nonumber\\
&-\left[\delta \check{\sigma}_{\rm imp}^{\sigma_\mu}, \check{g}_{\rm eq}^{\sigma_\mu}\right]+i{\bm v}_{\rm F}\cdot {\bm \nabla} \check{g}_{\rm eq}^{\sigma_\mu }=0,
\label{eq:transport_noneq}
\end{align} 
which is supplemented by $\{\check{g}_{\rm eq}^{\sigma_\mu }$, $\delta \check{g}^{\sigma_\mu }\}=0$ and the appropriate self-consistency conditions for the impurity self-energy.

We now implement this scheme for the thermal gradient. According to the discussion above, we need to make two steps. First, we assume a local equilibrium $T=T({\bm x})$ and expand the spatial gradient as ${\bm \nabla}\to {\bm \nabla}T\frac{\partial}{\partial T}+{\bm \partial}$, where ${\bm \partial}$ is the spatial gradient acting on the gap function~\cite{PhysRevB.53.15147,PhysRevB.75.224502}. 
Crucially, as discussed in the main text, the thermal gradient leads to changes in the pairing self-energy and we need to introduce a uniform supercurrent that maintains the Meissner state and compensates the thermoelectric quasiparticle current. Therefore, we second modify the gap function as $\Delta^{\sigma_\mu}_{\rm eq}({\bm k}_{\rm F})\to e^{i\bm Q \cdot \bm x} \Delta^{\sigma_\mu}_{\rm eq}({\bm k}_{\rm F})$, where $\bm Q$ is the center of mass momentum of Cooper pairs.
At this stage, $\bm Q$ is a parameter for calculations and should be determined to satisfy ${\bm J}_{\rm QP}+{\bm J}_{\rm SC}=0$~\cite{ginzburg1978thermoelectric,Shelly2016}. With this change, the gradient operates on the gap function and then Eq.~\eqref{eq:transport_noneq} becomes,
\begin{align}
&\left[\epsilon \check{\tau}_z- \check{\Delta}^{\sigma_\mu}_{\rm eq}({\bm k}_{\rm F})-\check{\sigma}^{\sigma_\mu}_{\rm imp, eq}, \delta \check{g}^{\sigma_\mu }\right]\nonumber\\
&-\left[\delta \check{\sigma}_{\rm imp}^{\sigma_\mu}, \check{g}_{\rm eq}^{\sigma_\mu}\right]+i{\bm v}_{\rm F}\cdot {\bm \nabla}T\frac{\partial}{\partial T}\check{g}_{\rm eq}^{\sigma_\mu}+i{\bm v}_{\rm F}\cdot {\bm \partial} \check{g}_{\rm eq}^{\sigma_\mu}=0.
\label{eq:transport}
\end{align} 
The third (fourth) term in Eq.~\eqref{eq:transport} describes the coupling between quasiparticles (Cooper pairs) and the thermal gradient.

It is convenient to separate the nonequilibrium QGF into the quasiparticle part, $\delta \check{g}_{\rm QP}^{\sigma_\mu}$, and the supercurrent part, $\delta \check{g}_{\rm SC}^{\sigma_\mu}$.
This separation divides Eq.~\eqref{eq:transport} into,
\begin{align}
\label{eq:transport_QP}
&\left[\epsilon \check{\tau}_z- \check{\Delta}^{\sigma_\mu}_{\rm eq}({\bm k}_{\rm F})-\check{\sigma}^{\sigma_\mu}_{\rm imp, eq}, \delta \check{g}_{\rm QP}^{\sigma_\mu }\right]\nonumber\\
&-\left[\delta \check{\sigma}_{\rm imp}^{\sigma_\mu}, \check{g}_{\rm eq}^{\sigma_\mu}\right]+i{\bm v}_{\rm F}\cdot {\bm \nabla}T\frac{\partial}{\partial T} \check{g}_{\rm eq}^{\sigma_\mu}=0,\\
\label{eq:transport_SC}
&\left[\epsilon \check{\tau}_z- \check{\Delta}^{\sigma_\mu}_{\rm eq}({\bm k}_{\rm F})-\check{\sigma}^{\sigma_\mu}_{\rm imp, eq}, \delta \check{g}_{\rm SC}^{\sigma_\mu }\right]+i{\bm v}_{\rm F}\cdot {\bm \partial} \check{g}_{\rm eq}^{\sigma_\mu}=0.
\end{align} 
Eq.~\eqref{eq:transport_QP} (Eq.~\eqref{eq:transport_SC}) describes the responses of quasiparticles (Cooper pairs) to the thermal gradient. 
In Eq.~\eqref{eq:transport_SC}, we set $\delta \check{\sigma}_{\rm imp}^{\sigma_\mu}=0$ because the supercurrent is not sensitive to impurities.

In principle, $\delta \check{g}_{\rm QP}^{\sigma_\mu}$ and $\delta \check{g}_{\rm SC}^{\sigma_\mu}$ are coupled with each other through the impurity self-energy and the normalization condition, $\{\check{g}^{\sigma_\mu}_{\rm eq}, \delta \check{g}^{\sigma_\mu}_{\rm QP}+\delta \check{g}^{\sigma_\mu}_{\rm SC} \}=0$.
However, note that if we solve Eqs.~(\ref{eq:transport_QP}-\ref{eq:transport_SC}) assuming that the anticommutators vanish individually, $\{\check{g}^{\sigma_\mu}_{\rm eq}, \delta \check{g}^{\sigma_\mu}_{\rm QP} \}=\{\check{g}^{\sigma_\mu}_{\rm eq}, \delta \check{g}^{\sigma_\mu}_{\rm SC} \}=0$, the resulting equation for the correction to the QGF, $\delta \varphi=\delta \check{g}^{\sigma_\mu}-\check{g}^{\sigma_\mu}_{\rm QP} -\check{g}^{\sigma_\mu}_{\rm SC}$, does not involve ``driving terms'' proportional to the thermal gradient. It follows that
in the linear response regime, the coupling between $\delta \check{g}_{\rm QP}^{\sigma_\mu}$ and $\delta \check{g}_{\rm SC}^{\sigma_\mu}$ introduces corrections only to nonlinear transport. Hence, we neglect it in our calculations. In the same approximation, the nonequilibrium part of the impurity self-energy is additive for the two components of quasiparticle and Cooper pairs and also satisfies the normalization condition (see Eq.~\eqref{impa} below).

\subsection*{Nonequilibrium quasiclassical Green function}
On the basis of the theoretical framework presented above, we derive the nonequilibrium QGF.
We first derive the expression for the quasiparticle contribution, $\delta \underline{g}_{\rm QP}^{\sigma_\mu{\rm X}}\;({\rm X=R,A})$, using Eq.~\eqref{eq:transport_QP}, which now reads,
\begin{align}
\left[\underline{M}^{\sigma_\mu {\rm X}}, \delta \underline{g}_{\rm QP}^{\sigma_\mu {\rm X}}\right]-\left[\delta \underline{\sigma}_{\rm imp}^{\sigma_\mu {\rm X}}, \underline{g}_{\rm eq}^{\sigma_\mu {\rm X}}\right]+i{\bm v}_{\rm F}\cdot {\bm \nabla}T\frac{\partial}{\partial T}\underline{g}_{\rm eq}^{\sigma_\mu {\rm X}}=0.
\label{eq:transport_QP_RA}
\end{align} 
Using the constraint $\{\underline{g}^{\sigma_\mu{\rm X}}_{\rm eq}, \delta \underline{g}^{\sigma_\mu{\rm X}}_{\rm QP} \}=0\;({\rm X=R,A})$ from Eq.~\eqref{eq:normalization}, we obtain,
\begin{align}
\label{delgRA}
\delta \underline{g}_{\rm QP}^{\sigma_\mu {\rm X}}=\frac{\underline{g}_{\rm eq}^{\sigma_\mu {\rm X}}}{2\pi D^{\sigma_\mu {\rm X}} } \left(\left[\delta \underline{\sigma}_{\rm imp}^{\sigma_\mu {\rm X}}, \underline{g}_{\rm eq}^{\sigma_\mu {\rm X}}\right]-i{\bm v}_{\rm F}\cdot {\bm \nabla}T\frac{\partial}{\partial T}\underline{g}_{\rm eq}^{\sigma_\mu {\rm X}}\right).
\end{align} 

We now turn to the Keldysh component, $\delta \underline{g}_{\rm QP}^{\sigma_\mu{\rm K}}$. Writing the corresponding matrix components of Eq.~\eqref{eq:transport_QP}, we have,
\begin{align}
\label{transport_gK}
&\left(\underline{M}^{\sigma_\mu {\rm R}} \delta \underline{g}_{\rm QP}^{\sigma_\mu {\rm K}}-\delta \underline{g}_{\rm QP}^{\sigma_\mu {\rm K}} \underline{M}^{\sigma_\mu {\rm A}}\right)
-\left(\sigma_{\rm imp,eq0}^{\sigma_\mu {\rm R}}-\sigma_{\rm imp,eq0}^{\sigma_\mu {\rm A}}\right)\delta \underline{g}_{\rm QP}^{\sigma_\mu {\rm K}}\nonumber\\
&+\left(\delta \underline{g}_{\rm QP}^{\sigma_\mu {\rm R}} \underline{\sigma}_{\rm imp,eq}^{\sigma_\mu {\rm K}}-\underline{\sigma}_{\rm imp,eq}^{\sigma_\mu {\rm K}}\delta \underline{g}_{\rm QP}^{\sigma_\mu {\rm A}}\right)\nonumber\\
&-\left(\delta \underline{\sigma}_{\rm imp}^{\sigma_\mu {\rm R}}\underline{g}^{\sigma_\mu{\rm K}}_{\rm eq}-\underline{g}^{\sigma_\mu{\rm K}}_{\rm eq}\delta \underline{\sigma}_{\rm imp}^{\sigma_\mu {\rm A}}\right)-\left(\delta \underline{\sigma}_{\rm imp}^{\sigma_\mu {\rm K}}\underline{g}^{\sigma_\mu{\rm A}}_{\rm eq}-\underline{g}^{\sigma_\mu{\rm R}}_{\rm eq}\delta \underline{\sigma}_{\rm imp}^{\sigma_\mu {\rm K}}\right)\nonumber\\
&+i{\bm v}_{\rm F}\cdot {\bm \nabla}T\frac{\partial}{\partial T}\underline{g}_{\rm eq}^{\sigma_\mu {\rm K}}=0,
\end{align}
which is supplemented by $\underline{g}^{\sigma_\mu{\rm R}}_{\rm eq} \delta  \underline{g}_{\rm QP}^{\sigma_\mu {\rm K}} +\delta  \underline{g}_{\rm QP}^{\sigma_\mu {\rm K}}\underline{g}^{\sigma_\mu{\rm A}}_{\rm eq}=0$.
Here, we defined for brevity $\sigma_{\rm imp,eq0}^{\sigma_\mu {\rm R,A}}={\rm Tr}(\underline{\sigma}_{\rm imp,eq}^{\sigma_\mu {\rm R,A}})$. 
To simplify Eq.~\eqref{transport_gK}, let us write $\delta \underline{g}_{\rm QP}^{\sigma_\mu{\rm K}}$ and $\delta \underline{\sigma}_{\rm imp}^{\sigma_\mu{\rm K}}$ as~\cite{PhysRevB.61.9061},
\beq
\label{ga}
\delta  \underline{g}_{\rm QP}^{\sigma_\mu {\rm K}}&=&
\left(\delta \underline{g}_{\rm QP}^{\sigma_\mu {\rm R}}-\delta \underline{g}_{\rm QP}^{\sigma_\mu {\rm A}} \right)\tanh \left(\frac{\epsilon}{2T}\right)+\delta  \underline{g}_{\rm QP}^{\sigma_\mu {\rm a}},\\
\label{impa1}
\delta  \underline{\sigma}_{\rm imp}^{\sigma_\mu {\rm K}}&=&
\left( \delta\underline{\sigma}_{\rm imp}^{\sigma_\mu {\rm R}}- \delta\underline{\sigma}_{\rm imp}^{\sigma_\mu {\rm A}} \right)\tanh \left(\frac{\epsilon}{2T}\right)+\delta  \underline{\sigma}_{\rm imp}^{\sigma_\mu {\rm a}}.
\eeq
The first term in Eq.~\eqref{ga} describes the changes in the spectral function while maintaining the distribution function in equilibrium.
The second term in Eq.~\eqref{ga} describes the changes in the distribution function and is essential for the response to the thermal gradient.
Using Eqs.~(\ref{ga}-\ref{impa1}), we recast Eq.~\eqref{transport_gK} as,
\begin{align}
\label{transport_ga}
&\left(\underline{M}^{\sigma_\mu {\rm R}} \delta \underline{g}_{\rm QP}^{\sigma_\mu {\rm a}}-\delta \underline{g}_{\rm QP}^{\sigma_\mu {\rm a}} \underline{M}^{\sigma_\mu {\rm A}}\right)
-\left(\sigma_{\rm imp,eq0}^{\sigma_\mu {\rm R}}-\sigma_{\rm imp,eq0}^{\sigma_\mu {\rm A}}\right)\delta \underline{g}_{\rm QP}^{\sigma_\mu {\rm a}}\nonumber\\
&+\left(\underline{g}^{\sigma_\mu{\rm R}}_{\rm eq}\delta \underline{\sigma}_{\rm imp}^{\sigma_\mu {\rm a}}-\delta \underline{\sigma}_{\rm imp}^{\sigma_\mu {\rm a}}\underline{g}^{\sigma_\mu{\rm A}}_{\rm eq}\right)\nonumber\\
&-\frac{i\epsilon{\bm v}_{\rm F}\cdot {\bm \nabla}T}{2T^2 \cosh^2 \left(\frac{\epsilon}{2T}\right)}
\left(\underline{g}^{\sigma_\mu{\rm R}}_{\rm eq}-\underline{g}^{\sigma_\mu{\rm A}}_{\rm eq}\right)
=0.
\end{align}
The corresponding constraint on $\delta  \underline{g}_{\rm QP}^{\sigma_\mu {\rm a}}$ from Eq.~\eqref{eq:normalization} becomes $\underline{g}^{\sigma_\mu{\rm R}}_{\rm eq} \delta  \underline{g}_{\rm QP}^{\sigma_\mu {\rm a}} +\delta  \underline{g}_{\rm QP}^{\sigma_\mu {\rm a}}\underline{g}^{\sigma_\mu{\rm A}}_{\rm eq}=0$.
Using Eqs.~(\ref{ga}-\ref{impa1}), we obtain the $T$-matrix equation for $\delta  \underline{\sigma}_{\rm imp}^{\sigma_\mu {\rm a}}$,
\begin{align}
\label{impa}
&\delta \underline{\sigma}^{\sigma_\mu {\rm a}}_{\rm imp}
=-\Gamma_{\rm imp} \left(\cot \delta +\left\langle \frac{\underline{g}^{\sigma_\mu {\rm R}}_{\rm eq}}{\pi} \right\rangle_{{\rm FS}} \right)^{-1}\nonumber\\
&\times \left\langle \frac{\delta \underline{g}_{\rm QP}^{\sigma_\mu {\rm a}}}{\pi} \right\rangle_{{\rm FS}}
 \left(\cot \delta +\left\langle \frac{\underline{g}^{\sigma_\mu {\rm A}}_{\rm eq}}{\pi} \right\rangle_{{\rm FS}} \right)^{-1}.
\end{align}
Using the constraint $\underline{g}^{\sigma_\mu{\rm R}}_{\rm eq} \delta  \underline{g}_{\rm QP}^{\sigma_\mu {\rm a}} +\delta  \underline{g}_{\rm QP}^{\sigma_\mu {\rm a}}\underline{g}^{\sigma_\mu{\rm A}}_{\rm eq}=0$ from Eq.~\eqref{eq:normalization}, we obtain,
\begin{subequations}
\beq
\label{delga}
\delta  \underline{g}_{\rm QP}^{\sigma_\mu {\rm a}}&=&\delta  \underline{g}^{\sigma_\mu {\rm a}}_{\rm ns}+\delta  \underline{g}^{\sigma_\mu {\rm a}}_{\rm vc},\\
\delta  \underline{g}^{\sigma_\mu {\rm a}}_{\rm ns}&=&\underline{N}_{\rm eq}^{\sigma_\mu{\rm R}}
\left(\underline{g}^{\sigma_\mu{\rm R}}_{\rm eq}-\underline{g}^{\sigma_\mu{\rm A}}_{\rm eq}\right)\left(\frac{-i\epsilon{\bm v}_{\rm F}\cdot {\bm \nabla}T}{2T^2 \cosh^2 \left(\frac{\epsilon}{2T}\right)}\right),\\
\delta  \underline{g}^{\sigma_\mu {\rm a}}_{\rm vc}&=&\underline{N}_{\rm eq}^{\sigma_\mu{\rm R}}\left(\underline{g}^{\sigma_\mu{\rm R}}_{\rm eq}\delta \underline{\sigma}_{\rm imp}^{\sigma_\mu {\rm a}}-\delta \underline{\sigma}_{\rm imp}^{\sigma_\mu {\rm a}}\underline{g}^{\sigma_\mu{\rm A}}_{\rm eq}\right),
\eeq
\end{subequations}
where
\beq
\underline{N}_{\rm eq}^{\sigma_\mu{\rm R}}=\frac{-\left(D^{\sigma_\mu {\rm R}}+D^{\sigma_\mu {\rm A}}\right)\frac{\underline{g}^{\sigma_\mu{\rm R}}_{\rm eq}}{\pi}+\sigma_{\rm imp,eq0}^{\sigma_\mu {\rm R}}-\sigma_{\rm imp,eq0}^{\sigma_\mu {\rm A}}}{\left(D^{\sigma_\mu {\rm R}}-D^{\sigma_\mu {\rm A}}\right)^2+\left(\sigma_{\rm imp,eq0}^{\sigma_\mu {\rm R}}-\sigma_{\rm imp,eq0}^{\sigma_\mu {\rm A}}\right)^2}.
\eeq
The first term in Eq.~\eqref{delga} only depends on the impurity self-energy in equilibrium.
This term describes the ``bare bubble diagram'': the spin current and the thermal current vertices connected by the two equilibrium QGFs in the diagrammatic calculations.
The second term involves the impurity self-energy in the nonequilibrium state and accounts for the vertex corrections~\cite{yip2016low}.
Indeed, expanding Eq.~\eqref{impa}, we 
find terms coexisting of both the advanced and retarded QGFs, such as $\epsilon {\bm v}_{\rm F}\underline{g}^{\sigma_\mu{\rm R}}_{\rm eq}\underline{g}^{\sigma_\mu{\rm A}}_{\rm eq}$.
Hence, $\delta \underline{\sigma}_{\rm imp}^{\sigma_\mu {\rm a}}$ contains the vertex corrections to the thermal current~\cite{mahan2013many}.
This second term is more essential for the SSSE and the SNE because the anisotropy between the electron and hole propagators appears through the impurity vertex corrections~\cite{matsushita2022spin}.

We finally consider $\delta \check{g}_{\rm SC}^{\sigma_\mu}$ using Eq.~\eqref{eq:transport_SC},
which takes the form,
\beq
\label{eq:transport_SC_RA}
\left[\underline{M}^{\sigma_\mu {\rm R,A}}, \delta \underline{g}^{\sigma_\mu {\rm R,A} }_{\rm SC} \right]+i {\bm v}_{\rm F}\cdot {\bm \partial} \underline{g}_{\rm eq}^{\sigma_\mu {\rm R,A}}=0\,. 
\eeq 
Using the constraint, $\{\underline{g}^{\sigma_\mu{\rm R,A}}_{\rm eq}, \delta \underline{g}^{\sigma_\mu{\rm R,A}}_{\rm SC} \}=0$ from Eq.~\eqref{eq:normalization}, we obtain, 
\begin{subequations}
\beq
\label{delgRA_SC}
\delta \underline{g}^{\sigma_\mu {\rm R,A} }_{\rm SC}&=&
\frac{\left({\bm v}_{\rm F}\cdot {\bm Q}\right) \underline{g}_{\rm eq}^{\sigma_\mu {\rm R,A}}\underline{\tau}_z \underline{\Delta}^{\sigma_\mu }_{\rm eq}}{2D^{\sigma_\mu {\rm R,A}2}},\\
\label{delgK_SC}
\delta \underline{g}^{\sigma_\mu {\rm K} }_{\rm SC}&=&
(\delta \underline{g}^{\sigma_\mu {\rm R} }_{\rm SC}-\delta \underline{g}^{\sigma_\mu {\rm A} }_{\rm SC})\tanh(\frac{\epsilon}{2T}).
\eeq 
\label{deltaG_SC}
\end{subequations}
We now have the nonequilibrium QGFs necessary to compute the spin current.

\subsection*{Spin current response to the thermal gradient}
In spin-triplet superconductors, both quasiparticles and Cooper pairs carry the spin and electric currents. 
With Eq.~\eqref{delgRA} and Eq.~\eqref{delga}, the spin and electric currents carried by the quasiparticles are expressed as~\cite{matsushita2022spin},
\begin{subequations}
\beq
\label{SC_QP}
{\bm J}_{\rm QP}^{\sigma_{\mu}}&=&\frac{N(\epsilon_{\rm F})}{2} \sum_{\sigma_{\mu}=\pm1} \int \frac{d\epsilon}{4\pi i} \left\langle{\rm Tr}\left[{\bm v_{\rm F}} \sigma_\mu \underline{\tau}_z  \delta \underline{g}^{\sigma_\mu {\rm K} }_{\rm QP}\right]\right\rangle_{{\rm FS}},\\
\label{EC_QP}
{\bm J}_{\rm QP}&=&\frac{N(\epsilon_{\rm F})}{2} \sum_{\sigma_{\mu}=\pm1}\int \frac{d\epsilon}{4\pi i} \left\langle{\rm Tr}\left[{\bm v_{\rm F}}  \underline{\tau}_z \delta \underline{g}^{\sigma_\mu {\rm K} }_{\rm QP}\right]\right\rangle_{{\rm FS}}.
\eeq
\end{subequations}
With Eqs.~\eqref{deltaG_SC}, the spin and electric supercurrents are expressed as,
\begin{subequations}
\beq
\label{SC_SC}
{\bm J}_{\rm SC}^{\sigma_{\mu}}&=&\frac{N(\epsilon_{\rm F})}{2} \sum_{\sigma_{\mu}=\pm1} \int \frac{d\epsilon}{4\pi i} \left\langle{\rm Tr}\left[{\bm v_{\rm F}} \sigma_\mu \underline{\tau}_z  \delta \underline{g}^{\sigma_\mu {\rm K} }_{\rm SC}\right]\right\rangle_{{\rm FS}},\\
\label{EC_SC}
{\bm J}_{\rm SC}&=&\frac{N(\epsilon_{\rm F})}{2} \sum_{\sigma_{\mu}=\pm1}\int \frac{d\epsilon}{4\pi i} \left\langle{\rm Tr}\left[{\bm v_{\rm F}}  \underline{\tau}_z \delta \underline{g}^{\sigma_\mu {\rm K} }_{\rm SC}\right]\right\rangle_{{\rm FS}}.
\eeq
\end{subequations}

We now determine the phase gradient, $\bm Q$, such that the thermoelectric current vanishes, ${\bm J}_{\rm QP}+{\bm J}_{\rm SC}=0$~\cite{ginzburg1978thermoelectric,Shelly2016}.
With Eq.~\eqref{EC_QP} and Eq.~\eqref{EC_SC}, this condition is expressed as,
\beq
\label{Q}
\sum_{\sigma_{\mu}=\pm1} \int \frac{d\epsilon}{4\pi i} \left\langle{\rm Tr}\left[{\bm v_{\rm F}} \sigma_\mu \underline{\tau}_z  \left(\delta \underline{g}^{\sigma_\mu {\rm K} }_{\rm QP}+\delta \underline{g}^{\sigma_\mu {\rm K} }_{\rm SC}\right)\right]\right\rangle_{{\rm FS}}=0\,.
\eeq
We solve Eq.~\eqref{Q} to obtain $\bm Q$, and then compute the spin current, ${\bm J}^{\sigma_{\mu}}={\bm J}_{\rm QP}^{\sigma_{\mu}}+{\bm J}_{\rm SC}^{\sigma_{\mu}}$, driven by the thermal gradient.

\subsection*{Acknowledgement}
T. Matsushita thanks M. Sato, Y. Yanase, T. Shibauchi, K. Hashimoto, K. Ishihara, J. Tei, and K. Shinada for fruitful discussions.
T. Matsushita appreciates Y. Araki, J. Fujimoto, and A. Ozawa for their discussion of the SNE. 
The authors thank D. Aoki and M. Shimizu for their discussion about the material parameters for UTe$_2$, and Y. Niimi for his discussion from an experimental point of view.
The numerical calculations were partially performed with computational facilities at YITP in Kyoto University. 

\subsection*{Funding}
T. Matsushita was supported by a Japan Society for the Promotion of Science (JSPS) Fellowship for Young Scientists, JSPS KAKENHI (Grant No.~JP19J20144, and No.~JP24KJ0130). I. V. was supported in part by grant NSF PHY-1748958 to the Kavli Institute for Theoretical Physics (KITP).
This work was also supported by JST CREST (Grant No.~JPMJCR19T2, and No.~JPMJCR19T5), and the Grant-in-Aid for Scientific Research on Innovative Areas ``Quantum Liquid Crystals (Grant No.~JP22H04480)'' from JSPS of Japan, and JSPS KAKENHI (Grants No. JP20K03860, No.~JP23K20828, No.~JP23K22492, and No.~JP24K17000). 

\subsection*{Availability} 
T. Matsushita and I. Vekhter conceived this work and constructed the linear response theory of the spin current to the thermal gradient in superconductors.
T. Matsushita and T. Mizushima performed the symmetry-based analysis for the SSSE and the SNE.
T. Matsushita performed all numerical calculations.
All authors discussed the calculated results and wrote the paper.

\subsection*{Competing interests} 
The authors declare that they have no competing interests. 

\subsection*{Availability} 
All data needed to evaluate the conclusions in the paper are presented in the paper {and/or Supplemental Materials.} 

\bibliographystyle{apsrev4-1_PRX_style}
\bibliography{SSE.bib}
\end{document}